\documentclass[floats,aps,amsfonts,notitlepage,nofootinbib]{revtex4-1}
\usepackage{ifpdf}
\usepackage[utf8]{inputenc}
\usepackage[UKenglish]{babel}
\usepackage{graphicx}
\usepackage{amsmath, amsthm, amssymb}
\usepackage{multirow}
\usepackage{url}
\usepackage{subfigure}
\usepackage{color}

\usepackage[colorlinks]{hyperref}   

\definecolor{CiteColor}{rgb}{0,0.5,0}
\hypersetup{citecolor=CiteColor}
\definecolor{RefColor}{rgb}{0.55,0,0}
\hypersetup{linkcolor=RefColor}
\definecolor{darkgreen}{rgb}{0.2,0.7,0.2}

\newcommand{\diff}[2]  {\frac{d #1}{d #2}}
\newcommand{\sdiff}[2]  {\frac{d^2 #1}{d #2^2}}
\newcommand{\pdiff}[2]  {\frac{\partial #1}{\partial #2}}

\newcommand{\cons}{{\text{cons}}}
\newcommand{\diss}{{\text{diss}}}
\newcommand{\self}{{\text{self}}}
\newcommand{\ret}{{\text{ret}}}
\newcommand{\en}{\mathcal{E}}
\newcommand{\ang}{\mathcal{L}_z}
\newcommand{\Carter}{\mathcal{Q}}

\renewcommand{\c}{\cos \omega_r \tau}
\renewcommand{\max}{\text{max}}
\renewcommand{\min}{\text{min}}

\newcommand{\be}{\begin{equation}}
\newcommand{\ee}{\end{equation}}
\newcommand{\ba}{\begin{eqnarray}}
\newcommand{\ea}{\end{eqnarray}}

\renewcommand{\ell}{{\hat{l}}}
\newcommand{\out}{\text{out}}
\renewcommand{\in}{\text{in}}
\newcommand{\degree}{^{\circ}}

\renewcommand{\c}{\hskip0.1cm,}
\newcommand{\p}{\hskip0.1cm.}

\mathchardef\mhyphen="2D

\def\prd{Phys. Rev. D}

\def\etal{\textit{et al.}}

\begin{document}

\title{Self force on a scalar charge in Kerr spacetime: inclined circular orbits}

\author{Niels Warburton}
\affiliation{School of Mathematical Sciences and Complex \& Adaptive Systems Laboratory, University College Dublin, Belfield, Dublin 4, Ireland}

\begin{abstract}
Accurately modeling astrophysical extreme-mass-ratio-insprials requires calculating the gravitational self-force for orbits in Kerr spacetime. The necessary calculation techniques are typically very complex and, consequently, toy scalar-field models are often developed in order to establish a particular calculational approach. To that end, I present a calculation of the scalar-field self-force for a particle moving on a (fixed) inclined circular geodesic of a background Kerr black hole. I make the calculation in the frequency-domain and demonstrate how to apply the mode-sum regularization procedure to all four components of the self-force. I present results for a number of strong-field orbits which can be used as benchmarks for emerging self-force calculation techniques in Kerr spacetime.
\end{abstract}	
\date{\today}
\maketitle

\section{Introduction}

The forthcoming advent of gravitational-wave astronomy necessitates accurate modeling of many astrophysical phenomena. Amongst the systems of interest are extreme-mass-ratio inspirals (EMRIs): binary systems where one of the components is substantially more massive than the other. The archetypal EMRI involves a stellar mass black hole or neutron star in orbit about a massive black hole, such as those now believed to exist at the center of most galaxies. Such systems are key sources for future space-based detectors and will allow the predictions of general relativity to be probed in the strong-field regime \cite{Gair:testing_GR}. EMRI systems are expected to undergo many thousands of orbits whilst emitting gravitational waves of frequencies observable by space-based detectors. The valuable information encoded in the waves will allow the spacetime of the massive black hole to be effectively `mapped out' \cite{Glampedakis-Babak:test_Kerr_hypothesis}. In particular, this will offer a resolution to the Kerr hypothesis: is the spacetime of an isolated astrophysical black hole described by the Kerr metric of general relativity?

Key to resolving such open questions will be accurate models of the gravitational waves emitted from EMRIs. Recently the leading-order dissipative dynamics of EMRI systems has been successfully modeled \cite{Mino,Drasco-Hughes,Sago-Tanaka-Hikida-Ganz-Nakano,Fujita-Hikida-Tagoshi}. In order to track the phase evolution over an entire inspiral using matched filtering techniques it will be necessary to go beyond the leading-order dissipative approximation and include both leading-order conservative and subleading-order dissipative corrections \cite{Hinderer-Flanagan,Gair:testing_GR}. The efforts of researchers to understand these corrections is usually known as the self-force program.

The standard self-force approach to studying EMRIs is to model the less massive body as a point particle and then calculate the orbital inspiral via the particle's interaction with its own metric perturbation. One of the main challenges of this approach is the need to regularize the divergent metric perturbation at the particle's location. This procedure is now well understood through to first-order-in-the-mass-ratio \cite{Mino-Sasaki-Tanaka,Quinn-Wald,Gralla-Wald,Poisson-review} and second-order-in-the-mass-ratio formulations are beginning to emerge \cite{Detweiler:2nd_order,Pound:2nd_order,Gralla:2nd_order,Pound:nonlinear_GSF,Pound-Miller,Pound:2nd_order_huu}.

Over the past decade or so, the goal of those working on self-force calculations has been to accurately model the motion of a compact object in orbit about a massive, rotating, Kerr black hole. Much progress has been made in this direction and there are now calculations that include conservative corrections to the dynamics of a compact object moving on generic orbits about non-rotating Schwarzschild black holes \cite{Barack-Sago-circular,Barack-Sago-eccentric,Detweiler-circular,Akcay-GSF-circular,Barack-Sago-ISCO-shift,Barack-Sago-precession,Warburton_etal,Akcay-Warburton-Barack,Shah-etal,Dolan_etal:spin_invariant,Dolan_etal:tidal_invariants}. Attention is now turning to extending this body of work to the most astrophysically relevant scenario of motion about a Kerr black hole, as well as including subleading-order dissipative corrections.

The recent progress with gravitational perturbations in Schwarzschild spacetime built heavily upon previous work involving a toy-model of a particle carrying a scalar charge \cite{Burko-circular,Canizares-Sopuerta:circular,Canizares-Sopuerta:eccentric,Haas,Detweiler-Messaritaki-Whiting,Diener_etal:SSF_self_consistent,Casals-Dolan-Ottewill-Wardell,Wardell_etal:Greens_func_Schwarz}. Progress on calculations in Kerr spacetime are proving to be no different. Two previous articles \cite{Warburton-Barack,Warburton-Barack:eccentric} (hereafter Papers I and II) presented the first calculations for a particle with scalar charge moving on circular, equatorial and eccentric, equatorial orbits about a Kerr black hole, respectively. The results of these works have since been used as benchmarks for alternative approaches to the scalar-field problem \cite{Dolan-Barack,Thornburg-Wardell}, one of which has since been extended to the Lorenz-gauge gravitational case for circular, equatorial orbits in Kerr geometry \cite{Dolan-Barack:GSF_Schwarz,Dolan-Barack:GSF_Kerr}. There has also been progress calculating the gravitational self-force in radiation gauges \cite{Shah-etal,Shah_etal:Kerr,Pound-Merlin-Barack,Merlin-Shah:2014}. Knowledge of the gravitational self-force, in both gauges, has recently been employed to compute the gauge-invariant shift in the frequency of the inner-most stable circular equatorial orbit about a Kerr black hole \cite{Isoyama_etal}.

The aim of this work is to extend Papers I and II to provide further benchmarks for emerging self-force calculation techniques. In particular I will, for the first time, present results for the scalar-field self-force (SSF) for orbits not confined to the equatorial plane. In this work, as well as Papers I and II, I solve the scalar-field wave equation coupled to a point-particle source. In making these calculations I have opted to work in the frequency-domain for reasons I outline now.

In Schwarzschild spacetime the angular dependence of the scalar-field wave equation can be separated by decomposing the field and source into spherical-harmonic modes. One then has the choice to solve the resulting 1+1D set of partial differential equations (PDEs) in the time-domain or to further decompose into Fourier modes and solve the resulting ordinary differential equations (ODEs) for the radial component of the field. Time domain self-force calculations in Kerr spacetime usually involve solving the full 3+1D or 2+1D field equations. Within these decompositions the retarded field is formally divergent at the particle's location and so effective-source techniques are employed to regularize the field \cite{Vega-Detweiler,Barack-Golbourn-Sago,Vega-Wardell-Diener,Diener_etal:SSF_self_consistent,Warburton-Wardell}. A 1+1D decomposition into spherical-harmonic modes is also possible in Kerr spacetime and this has the attraction that each multipole mode of the scalar-field is finite at the particle's location. The downside to this approach is that, in Kerr spacetime, the resulting field equations couple between the individual multipole modes (though recently it has been shown that this coupling is tractable in practice \cite{Dolan:Kerr_1+1}). 

For a complete separation of the angular dependence of the scalar-field equation in Kerr spacetime it is necessary to decompose into both spheroidal-harmonic and Fourier modes \cite{Carter, Brill}. Then, as with a spherical-harmonic decomposition, the individual modes of the scalar-field are finite at the particle but now the individual multipole modes also decouple from one-another. This allows each multipole mode to be solved for in isolation from all the others. Working in the frequency-domain is also attractive because one only encounters ODEs which are numerically much more straightforward to solve than PDEs. One of the goals of this work is to provide benchmarks for other emerging self-force calculation techniques and the ease with which ODEs can be solved to high accuracy greatly assists in achieving this aim. It is for these reasons that in this work I have chosen to pursue a frequency-domain approach.

The layout of this paper is as follows. Section \ref{sec:setup} details the orbital parameterization for inclined circular orbits and gives the equations for scalar-field perturbations in Kerr spacetime. Section \ref{sec:SF_via_mode_sum} outlines the mode-sum approach and shows how to apply it to all four components of the scalar-field self-force. Section \ref{sec:numerical_implementation} overviews my numerical implementation with results presented in Sec.~\ref{sec:results}. Throughout this work I use Boyer-Lindquist coordinates $(t,r,\theta,\varphi)$ with metric signature $(-+++)$ and geometric units such that the speed of light and the gravitational constant are equal to unity.

\section{Equations of motion and perturbation formalism}\label{sec:setup}

In this work I shall consider the SSF experienced by a particle moving on an inclined, circular geodesic of fixed Boyer-Lindquist radius in Kerr spacetime, ignoring back-reaction effects on the orbit. I shall denote the particle's worldline by $x^\mu_p(\tau)$ and its four-velocity by $u^\mu(\tau) = dx_p^\mu/d\tau$ where $\tau$ is the particle's propertime and hereafter a subscript `$p$' is used to denote a coordinate's value at the particle. I shall use $M$ and $aM$ to denote the black hole's mass and spin, respectively. In this work I break with the convention from Papers I and II and take $a\ge0$, instead letting the sign of the orbital angular momentum, $\ang$, differentiate between prograde and retrograde motion ($\ang>0$ prograde, $\ang<0$ retrograde). This convention allows for smoothly varying the orbital inclination (defined below) from prograde orbits to retrograde orbits without flipping the sign of $a$. 

I now briefly consider the generic motion of test particles about a Kerr black hole before specializing to inclined circular orbits. It is well know that the equations for geodesic motion in Kerr spacetime can be expressed in first-order form \cite{Carter}. When the equations are written this way one encounters three constants of motion: the specific energy $\en = -u_t$, the specific azimuthal angular-momentum $\ang = u_\varphi$, and the Carter constant $\mathcal{Q}$. The former two are related to the Killing vectors of the background spacetime, and the latter relates to a Killing tensor. The explicit first-order form of the equations of motion for a timelike test body in Kerr spacetime are given by \cite{MTW}
\begin{eqnarray}
	\rho^4 	\left ( \diff{r}{\tau} 		\right)^2 	&=& \left[ \en(r^2+a^2) - a\ang \right]^2 - \Delta \left[ r^2 + (\ang - a\en)^2 + \mathcal{Q} \right]		\equiv V_r			\c		\label{eq:V_r}	\\
	\rho^4	\left ( \diff{\theta}{\tau} \right)^2 	&=&	\mathcal{Q} - \cot^2 \theta \ang^2 - a^2 \cos^2\theta(1-\en)^2										\equiv V_\theta		\c		\label{eq:V_theta}	\\
	\rho^2	\left ( \diff{\varphi}{\tau}	\right)	&=&	\csc^2\theta \ang + a \en \left( \frac{r^2 + a^2}{\Delta} - 1 \right) - \frac{a^2 \ang}{\Delta}					\c		\label{eq:V_phi}	\\
	\rho^2	\left ( \diff{t}{\tau}		\right)	&=& \en \left[ \frac{(r^2 + a^2)^2}{\Delta} - a^2\sin^2\theta \right] + a \ang \left( 1 - \frac{r^2 + a^2}{\Delta} \right)	\label{eq:V_t} 	\c
\end{eqnarray}
where
\begin{eqnarray}
	\Delta \equiv r^2 - 2Mr + a^2 \c \qquad \rho^2 \equiv r^2 + a^2\cos^2\theta 	\p
\end{eqnarray}
For generic orbits the roots of $V_r$ and $V_\theta$ give the radial and polar orbital turning points, respectively. In the following subsections I give a useful parametrization for the case of inclined circular orbits and then discuss the decomposition of the scalar-field wave equation into the frequency-domain.

\subsection{Orbital parametrization}\label{sec:orbit_param}

For a given black hole spin, the family of inclined circular orbits can be parametrized by the pair $(r_0,\iota$) where $r_0$ is the Boyer-Lindquist radius of the orbit and the inclination angle, $\iota$, is related to the fundamental constants through
\begin{equation}
	\cos \iota = \frac{\ang}{\sqrt{\ang^2 + \mathcal{Q}}}\p
\end{equation}

In order to avoid divergences at the orbital turning points, involving terms such as $(d\theta/d\tau)^{-1}$, it is convenient to transform to a new set of coordinates to describe the orbit. Hughes provides one such parameterization \cite{Hughes} which I give now for completeness. Defining $z=\cos^2\theta_p$, Eq.~\eqref{eq:V_theta} becomes
\begin{eqnarray}
	\diff{\theta_p}{\tau} &=& \pm \frac{ \sqrt{[a^2(1-\en^2)] z^2 - [\mathcal{Q}+\ang^2+a^2(1-\en^2)] z + \mathcal{Q}}}{(r_0^2+a^2z)\sqrt{1-z}}\c		\label{eq:dtheta_dt}	\\
						&=&	\pm \frac{ \sqrt{\beta(z_+ - z)(z_- - z)} } {(r_0^2 + a^2 z)\sqrt{1-z}}\c
\end{eqnarray}
where $\beta \equiv a^2(1-\en^2)$ and $z_\pm$ are the two roots of the quadratic equation in the numerator of Eq.~\eqref{eq:dtheta_dt}. The upper sign corresponds to motion from $\theta_\text{min}$ to $\theta_\text{max}$ and vice versa for the lower sign.

Further defining $z=z_- \cos^2\chi$, where $\chi$ is a monotonically increasing parameter along the particle's worldline with $\theta_p=\theta_\text{min}$ at $\chi=0,2\pi\dots$ and $\theta_p=\theta_\text{max}$ at $\chi=\pi,3\pi\dots$, gives
\begin{eqnarray}
	\diff{\chi}{\theta_p} = \frac{d\chi/dz}{dz/d\theta_p} = 	\pm \sqrt{\frac{1-z}{z_- - z}} \c				\label{eq:dchi_dtheta}																	
\end{eqnarray}
where the $\pm$ has the same meaning as in Eq.~\eqref{eq:dtheta_dt}. The polar angle as a function of $\chi$ is then computed via
\begin{equation}
	\theta_p(\chi) = \theta_{\min} + \int^\chi_0 \diff{\theta_p}{\chi'}\,d\chi',\qquad \theta_{\min} = \cos^{-1}(\sqrt{z_-})\p
\end{equation}
Combining Eqs.~\eqref{eq:dtheta_dt} and \eqref{eq:dchi_dtheta} gives
\begin{eqnarray}
	\diff{\chi}{\tau} = \frac{ \sqrt{\beta(z_+ - z)} }{r_0^2+a^2z}	\p		\label{eq:dchi_dtau}
\end{eqnarray}
Further combining Eq.~\eqref{eq:dchi_dtau} with Eqs.~\eqref{eq:V_phi} and \eqref{eq:V_t} gives
\begin{eqnarray}
	\diff{t_p}{\chi} = \frac{ \gamma + a^2 \en z}{\sqrt{\beta(z_+ - z)}} \c		\qquad		\diff{\varphi_p}{\chi} = \frac{1}{\sqrt{\beta (z_+ - z)}} \left( \frac{\ang}{1-z} + \delta \right)\c
\end{eqnarray}
with
\begin{eqnarray}
	\gamma = \en \left[ \frac{(r_0^2 + a^2)^2}{\Delta} - a^2 \right] + a \ang \left( 1 - \frac{r_0^2 + a^2}{\Delta} \right) \c \qquad \delta = a\en \left( \frac{r_0^2 + a^2}{\Delta} - 1 \right) - \frac{a^2\ang}{\Delta}\p
\end{eqnarray}
Lastly, $t_p$ and $\varphi_p$ as functions of $\chi$ are given by
\begin{eqnarray}
	t_p(\chi)		= \int_0^\chi \diff{t}{\chi'} d\chi'			\c \qquad	\varphi_p(\chi)	=		\int_0^\chi \diff{\varphi}{\chi'} d\chi'\c
\end{eqnarray}
where I have assumed the initial periastron passage occurs at $t_p=\varphi_p=0$.

The constants of the orbital motion that appear above can be written in terms of $r_0$ and $\ang$ in the following manner. For a black hole of spin $a$ and a given orbit with radius $r_0$ and angular momentum $\ang$, solving $V_r = d V_r/ d\tau = 0$ gives $\en$ and $\mathcal{Q}$ as
\begin{eqnarray}
	\en(r_0,\ang) 	&=&		\frac{a^2 \ang^2(r_0-M) + r_0\Delta_0^2}{ a\ang M(r_0^2 - a^2) \pm \Delta_0 \sqrt{ r_0^5(r_0-3M) + a^4r_0(r_0+M) + a^2r_0^2(\ang^2 - 2Mr_0 + 2r_0^2) } }		\label{eq:energy}		\c\hspace{8mm}		\\
	\mathcal{Q}(r_0,\ang)	&=&		\frac{\left[(a^2+r_0^2) \en(r_0,\ang) - a\ang\right]^2}{\Delta_0} - \left[ r_0^2 + a^2 \en(r_0,\ang)^2 - 2a \en(r_0,\ang)\ang + \ang^2 \right]\p
\end{eqnarray}
where $\Delta_0 = \Delta(r_0)$. Of the two roots in Eq.~\eqref{eq:energy} it turns out that the minus sign is only relevant for nearly horizon-skimming orbits about rapidly rotating black holes \cite{Hughes:horizon_skimming}. In this work I will not consider such orbits and so will always take the plus sign.

Let the Boyer-Lindquist time, $t$, taken for the particle to complete one orbit (i.e., the time taken for $\theta_p$ to go from $\theta_\min$ to $\theta_\max$ and back again) be denoted by $T_\theta \equiv t_p(2\pi) = 2t_p(\pi)$. The azimuthal angle swept out during this time I will denote by $\Delta\varphi_p \equiv \varphi_p(2\pi) = 2\varphi_p(\pi)$. Using $T_\theta$ and $\Delta\varphi_p$ the polar and azimuthal orbital frequencies are given by
\begin{eqnarray}\label{eq:circ-inclined-freqs}
	\Omega_\theta = \frac{2\pi}{T_\theta}	\c \qquad 	\Omega_\varphi = \frac{\Delta \varphi_p}{T_\theta}\p
\end{eqnarray}

\subsubsection{Schwarzschild limit}\label{sec:orbit_param_Schwarz}

Later, as a test on my numerical code, I will present results for the SSF along an inclined circular orbit in Schwarzschild spacetime. The above orbital parameterization is ill-defined when directly setting $a=0$ as, for example, $\beta^{-1}$ and $z_+$ diverge as $a\rightarrow0$. By carefully taking the limit to $a\rightarrow0$, the required equations for an inclined circular orbit in Schwarzschild spacetime are given by
\begin{align}
	z_-(a=0) &= 1 -\frac{r_0-3M}{r_0^2 M} \ang^2,\qquad\left.\frac{dt}{d\chi}\right|_{a=0} = \left(\frac{r_0^3}{M}\right)^{1/2},										\label{eq:zminus_a_0}\\
	\left.\frac{d\varphi}{d\chi}\right|_{a=0} &= \frac{2\ang r_0(r_0-3M)^{1/2}M^{1/2}}{\ang^2(r-3M)+r_0^2 M +\left[\ang^2(r_0-3M)-r_0^2 M\right]\cos(2\chi)}\p
\end{align}
All the other orbital parameterization equations are well defined when setting $a=0$ so long as $z_-$ is replaced by $z_-(a=0)$ from Eq.~\eqref{eq:zminus_a_0}. Also note that for inclined circular orbits in Schwarzschild spacetime the orbital frequencies are degenerate, i.e., $\Omega_\theta = \Omega_\varphi$.

\subsection{Perturbation formalism and multipole decomposition}\label{sec:scalar_field_decomp}

In this work I shall consider the particle to be carrying a scalar charge, $q$. The scalar field that arises from this charge I shall take to be governed by the minimally coupled Klein-Gordon equation:
\begin{eqnarray}\label{eq:fieldeq}
	\square \Phi \equiv \nabla_\alpha \nabla^\alpha \Phi = -4\pi T \c
\end{eqnarray}
where $\nabla_\alpha$ represents covariant differentiation with respect to the background Kerr metric and $T$ denotes the particle's scalar charge density. In a given coordinate system the D'Alembertian operator can be expressed as
\begin{eqnarray}
	\square \Phi = [-\det(g)]^{-1/2} \pdiff{}{x^\mu} \left( g^{\mu\nu}\left[-\det(g)\right]^{1/2} \pdiff{\Phi}{x^\nu} \right)\c
\end{eqnarray}
where $g$ is the background Kerr metric and $\det(g)$ is the metric determinant with $\det(g) = - \rho^4 \sin^2 \theta$ in Boyer-Lindquist coordinates. In this work the scalar charge density will be a $\delta$-function along the particle's world line:
\begin{eqnarray}\label{eq:T_particle}
	T				 	&=& q \int \delta^4 ( x^\mu - x_p^\mu(\tau) ) [-g(x)^{-1/2}]  d\tau			\c						\label{eq:source}	\\
						&=&	\frac{q}{\rho^2\sin \theta u^t} \delta(r-r_p) \delta(\theta - \theta_p) \delta(\varphi - \varphi_p)	\c	\nonumber
\end{eqnarray}
where the second equation is obtained by changing integration variable from $\tau$ to $t$ in the first equation. Note that the $t$-component of the four-velocity $u^t$ is simply calculated as $u^t = g^{t\varphi} \ang - g^{tt}\en$. 

As discussed in the introduction, the scalar wave equation \eqref{eq:fieldeq} in Kerr geometry can be completely separated into spheroidal-harmonic and frequency modes in the form \cite{Brill}
\begin{eqnarray}\label{eq:field-decomp}
    \Phi = \int\sum_{\ell=0}^\infty\sum_{m=-\ell}^\ell R_{\ell m\omega}(r)S_{\ell m}(\theta;\sigma^2) e^{im\varphi} e^{-i\omega t}\, d\omega   \p
\end{eqnarray}
Here $S_{\ell m}(\theta;\sigma^2)$ are spheroidal Legendre functions with \textit{spheroidicity} $\sigma^2$ [I reserve the term {\it spheroidal harmonic} for the product $S_{\ell m}(\theta;\sigma^2)e^{im\varphi}$]. Notice that I label spheroidal-harmonic modes by $\ell m$, as I will later introduce {\em spherical}-harmonic modes which will be labelled by $lm$.  The spheroidal harmonics I use are orthonormal with normalization given by
\begin{eqnarray}\label{eq:spheroidal-harmonic-normalization}
    \oint S_{\ell m}(\theta;\sigma^2)e^{im\varphi}S_{\ell' m'}(\theta;\sigma^2)e^{-im'\varphi} d\Omega = \delta_{\ell\ell'} \delta_{mm'} \c 
\end{eqnarray}
with area element $d\Omega=\sin\theta d\theta d\varphi$, and with $\delta_{n_1n_2}$ being the standard Kronecker delta. 

The source spectra for inclined circular orbits is given by \cite{Hughes}
\begin{eqnarray}\label{eq:omega}
	\omega \equiv \omega_{mk} = 	m\Omega_\varphi + k\Omega_\theta\c		
\end{eqnarray} 
where $m$ and $k$ are integers and $\Omega_\varphi$, $\Omega_\theta$ are given in Eqs.~\eqref{eq:circ-inclined-freqs}. The nature of the source spectra implies that the integral in Eq.~\eqref{eq:field-decomp} can be rewritten as a discrete sum over  Fourier modes. The point particle source is decomposed in a similar fashion to the field as
\begin{eqnarray}\label{eq:source-decomp}
     \rho^2 T = \sum_{\ell=0}^\infty\sum_{m=-\ell}^\ell \sum_{k=-\infty}^{\infty} \tilde{T}_{\ell mk}(r)S_{\ell m}(\theta;\sigma^2) e^{im\varphi} e^{-i\omega_{mk} t}  \c
\end{eqnarray}
where the $\rho^2$ factor is introduced for later convenience. Using the orthonormal properties  \eqref{eq:spheroidal-harmonic-normalization} of spheroidal harmonics and taking the inverse Fourier transform of \eqref{eq:source-decomp}, the radial dependence of the source is found to be
\begin{eqnarray}\label{eq:T_FD}
	 \tilde{T}_{\ell mk}(r) = \frac{q}{T_\theta} \int^{T_\theta}_0	\frac{ S_{\ell m} (\theta_p(t);-a^2\omega_{mk}^2)}{u^t(r_0,\theta_p(t))}e^{i(\omega_{mk} t-m\varphi_p(t))}\delta(r-r_0) dt	\c
\end{eqnarray}
For circular equatorial orbits ($r_p=r_0$, $\varphi_p = \Omega_\varphi t$, $\omega \equiv \omega_m = m\Omega_\varphi$, $\theta_p=\pi/2$) the above equation reduces to
\begin{eqnarray}
	\tilde{T}_{\ell m}(r_0) = q \frac{S_{\ell m} (\pi/2;-a^2\omega_m^2)}{u^t(r_0,\pi/2)}\delta(r-r_0)\c \qquad\text{circular equatorial}\p
\end{eqnarray}
For inclined circular orbits I use the integers $\ell,$ $m$ and $k$ to index each mode of the scalar-field. Note there is no sum over the polar index, $k$, in the circular equatorial case \cite{Warburton-Barack}.

Substituting the field decomposition \eqref{eq:field-decomp} into the field equation (\ref{eq:fieldeq}) and using the source decomposition above, the radial and angular equations are found to be
\begin{align}
    &\Delta\pdiff{}{r}\left(\Delta\pdiff{R_{\ell mk}}{r}\right) + \left[a^2m^2-4Mrma\omega_{mk} + (r^2+a^2)^2\omega_{mk}^2 -a^2\omega_{mk}^2\Delta - \lambda_{\ell m}\Delta)\right]R_{\ell m k} \nonumber \\
	&= -4\pi\Delta_0  \tilde{T}_{\ell mk}(r)    \label{eq:radialeq} \c
\end{align}
\begin{equation}
    \frac{1}{\sin\theta}\pdiff{}{\theta}\left(\sin\theta\pdiff{S_{\ell mk}}{\theta}\right) + \left(\lambda_{\ell m} + a^2\omega_{mk}^2\cos^2\theta - \frac{m^2}{\sin^2\theta}\right)S_{\ell mk} = 0 \c \label{eq:angulareq}
\end{equation}
where, recall, $\Delta\equiv r^2-2Mr+a^2$ and $\Delta_0\equiv\Delta(r_0)$ and I have defined $R_{\ell mk}\equiv R_{\ell m\omega_{mk}}$ and $S_{\ell mk}\equiv S_{\ell m}(\theta;-a^2\omega^2_{mk})$. The angular equation (\ref{eq:angulareq}) takes the form of the spheroidal Legendre equation with spheroidicity $\sigma^2=-a^2\omega_{mk}^2$. Its eigenfunctions are the spheroidal Legendre functions $S_{\ell m}(\theta;-a^2\omega_{mk}^2)$ and its eigenvalues are denoted by $\lambda_{\ell m}$. In general there is no closed form for $S_{\ell m}$ or $\lambda_{\ell m}$, but they can be calculated using the spherical harmonic decomposition method described in Paper I.  When $a=0$, the spheroidal harmonics $S_{\ell m}e^{i m \varphi}$ coincide with the spherical harmonics $Y_{\hat lm}$ and their eigenvalues reduce to $\lambda_{\ell m} =\ell(\ell+1)$.

To further simply the field equations it is convenient to transform to a new variable
\begin{eqnarray}
    \psi_{\ell mk}(r)\equiv rR_{\ell mk}(r)       \c          \label{eq:R-psi}
\end{eqnarray}
and introducing a tortoise radial coordinate $r_*$ defined through
\begin{equation}\label{eq:rs}
    \diff{r_*}{r} = \frac{r^2}{\Delta} \p
\end{equation}
With the above definition the tortoise coordinate is given explicitly in terms of $r$ as
\begin{eqnarray}\label{eq:tortoise-explicit}
    r_* = r + M \ln(\Delta/M^2) + \frac{(2M^2 - a^2)}{2(M^2-a^2)^{1/2}}\ln\left(\frac{r-r_+}{r-r_-}\right)  \c
\end{eqnarray}
where the constant of integration has been specified and $r_\pm=M\pm\sqrt{M^2-a^2}$. 

In terms of $\psi_{\ell mk}(r)$ and $r_*$, the radial equation (\ref{eq:radialeq}) takes the simpler form,
\begin{eqnarray}\label{eq:rsradialeqn}
   \sdiff{\psi_{\ell mk}}{r_*} + W_{\ell mk}(r)\psi_{\ell m k} =  -\frac{4 \pi\Delta_0}{r_0^3} \tilde{T}_{\ell mk}  \c
\end{eqnarray}
where $\tilde{T}_{\ell mk}$ is given in Eq.~\eqref{eq:T_FD} and $W_{\ell mk}$ is an effective radial potential given by
\begin{eqnarray}
    W_{\ell mk}(r) = \left[\frac{(r^2+a^2)\omega_{mk} - am}{r^2}\right]^2 - \frac{\Delta}{r^4}\left[\lambda_{\ell m} - 2am\omega_{mk} + a^2\omega_{mk}^2 + \frac{2(Mr-a^2)}{r^2}\right]\p
\end{eqnarray}
There is no known closed-form, analytic solution to the radial equation \eqref{eq:rsradialeqn} for general $\ell m k$ and thus I opt to solve it numerically (a popular alternative approach would be to solve the homogeneous radial equation as as series expansion of special functions \cite{MST,MST_review}). The details of my numerical procedure are given below in Sec.~\ref{sec:numerical_implementation}.

\subsection{Self-force equations of motion}\label{sec:equations_of_motion}

In this section I outline the equations of motion for a particle coupled to a scalar-field \cite{Quinn}. As there are no known fundamental scalar fields of the type considered in this work there is a wide scrope for choosing the force law (the only known fundamental scalar field is that of the Higgs boson which is a complex, self-interacting scalar field). Here I take perhaps the simplest choice for the force law and discuss a curious consequence of this decision. With $\Phi^R$ as the smooth Detweiler-Whiting regular field \cite{Detweiler-Whiting} a common choice for the force law is
\begin{equation}\label{eq:force_law}
	u^\beta\nabla_\beta(\mu u^\alpha) = q \nabla^\alpha \Phi^R(x_p) \equiv F^\alpha_\self(x_p)\p
\end{equation}
Precisely how to construct $\Phi^R$ in practice will be discussed in the following section. An interesting feature of Eq.~\eqref{eq:force_law} is that the resulting SSF has a component tangential to the particle's four-velocity so that $u_\alpha F^\alpha_\self$ is generally non-zero. The consequence of this can be seen by expanding the derivative in Eq.~\eqref{eq:force_law}, whereupon one finds a term orthogonal to the particle's four-velocity, which is responsible for driving the orbital dynamics, and a term tangential to the four-velocity, which gives rise to a dynamically varying rest mass. The expanded equations read
\begin{eqnarray}
\mu\diff{u^\alpha}{\tau} 	&=& (\delta^\alpha_\beta + u^\alpha u_\beta)F^\beta_{\text{self}} 	\equiv F^\alpha_{\perp(\text{self})}	\c \label{eq:orthogonal-force}	\\
\diff{\mu}{\tau} 			&=& - u^\alpha F_\alpha^{\text{self}} \p				\label{eq:mass-change}
\end{eqnarray}
By combining Eqs.~\eqref{eq:force_law} and \eqref{eq:mass-change} the rest mass can be written explicitly as a function of $\tau$:
\begin{eqnarray}\label{eq:mass-change-explicit}
	\mu(\tau) = \mu_0 - q \Phi^R(\tau)	\c
\end{eqnarray}
where $\mu_0$ is a constant of integration (sometimes called the bare mass). For a stationary setup (where $\Phi^R$ is constant in time) the rest mass of the particle will remain constant along the orbit, but for more general setups the rest mass will vary with $\tau$. This unusual feature of this particular scalar-field setup can be understood `physically' by noting that a scalar charge can radiate monopole waves with the radiated energy coming at the expense of the particle's rest mass \cite{Poisson-review}. It turns out to be possible to construct a scalar field theory where the rest mass is conserved, but at the cost of the linearity of the resulting theory \cite{Quinn}. For this reason I choose to work with a scalar field governed by the Klein-Gordon equation, even though the resulting theory has a time-dependent rest mass.

Lastly, I note that in the setup outlined above the scalar charge is not necessarily conservered. For simplicity, I shall assume that $q$ remains constant as is commonly done by other authors \cite{Burko-Harte-Poisson, Haas-Poisson-mass_change}.

\section{Self-force via mode-sum regularization}\label{sec:SF_via_mode_sum}

Building on the work of Mino, Sasaki and Tanaka \cite{Mino-Sasaki-Tanaka} and Quinn and Wald \cite{Quinn-Wald}, Detweiler and Whiting demonstrated that the self-force can be computed as the derivative of a suitable regular field at the particle  --- see Eq.~\eqref{eq:force_law}. Formally the regular field, $\Phi^R$, is constructed by taking the standard retarded solution to Eq.~\eqref{eq:fieldeq}, which I denote by $\Phi^\ret$, and subtracting the appropriate singular component of the field, which I denote by $\Phi^S$. Both $\Phi^\ret$ and $\Phi^S$ are solutions to the sourced wave equation \eqref{eq:fieldeq} and, as a consequence, their difference is a solution to the homogeneous wave equation:
\begin{equation}
	\square \Phi^{\ret/S} = -4\pi T\,, \qquad \square \Phi^R = \square(\Phi^\ret - \Phi^S) = 0\p
\end{equation}
Formally the self-force can be calculated via
\begin{equation}\label{eq:F_self_as_difference}
	F_\alpha^\self(x_p) \equiv  q\nabla_\alpha \Phi^R (x_p) = q \lim_{x\rightarrow x_p}  \nabla_\alpha \left[ \Phi^\ret(x) - \Phi^S(x) \right] = \lim_{x\rightarrow x_p} \left[ F_\alpha^\ret(x) - F_\alpha^S(x) \right]	 \c
\end{equation}
where $F_\alpha^{\ret/S}(x) \equiv q\nabla_\alpha \Phi^{\ret/S}(x)$.

Equation \eqref{eq:F_self_as_difference} is not in a practical form, as both $F^\alpha_\ret(x)$ and $F^\alpha_S(x)$ diverge in the limit $x\rightarrow x_p$. A more practical approach is the mode-sum prescription, whereby the full retarded field, regular field and the singular field are decomposed into {\it scalar} spherical-harmonic modes. This decomposition has the advantage that the individual $lm$-modes of the retarded and singular fields are finite at the particle's location. Within the mode-sum approach the force due to the regular field is written as
\begin{equation}
	F_\alpha^\self(x_p) = \lim_{x\rightarrow x_p} \sum_l \left[ F_{\alpha}^{(\ret)l}(x) - F_{\alpha}^{(S)l} (x)\right]		\c
\end{equation}
where $F_{\alpha}^{(\ret/S)l}$ denotes the spherical-harmonic $l$-mode contribution (summed over $m$) to $F_{\alpha}^{\ret/S}$. Generally the retarded force per $l$-mode has to be computed numerically and I present the details of this calculation in Sec.~\ref{sec:numerical_implementation} below. The singular piece on the other hand is accessible to an analytical treatment. The structure of the singular component of the field was first analyzed by Mino \etal~\cite{Mino-Sasaki-Tanaka} and the practical mode-sum method for computing the SF was developed shortly after by Barack and Ori \cite{mode-sum-orig,Barack-Ori}. The formula they obtained for regularizing the force is given by
\begin{eqnarray}\label{eq:ret-mode-sum}
	F^{\text{self}}_\alpha(x_p) = q \sum_{l=0}^\infty \left( F^{(\ret)l}_{\alpha \pm}(x_p) - A_{\alpha\pm} L - B_\alpha - C_\alpha L^{-1} \right) \c
\end{eqnarray}
where $L=l+1/2$. Each $F_{\alpha}^{(\ret)l}$ is finite at the particle’s location, although in general the sided limits $r \rightarrow r_p^\pm$ yield two different values, denoted $F_{\alpha\pm}^{(\ret)l}$ respectively. The coefficients $A_\alpha, B_\alpha, C_\alpha$ are $l$-independent {\it regularization parameters}, the values of which are known for generic bound orbits about a Schwarzschild \cite{mode-sum-orig} or Kerr black hole \cite{Barack-Ori}. 

As the series in Eq.\ \eqref{eq:ret-mode-sum} is truncated at $\mathcal{O}(L^{-1})$ it is expected, for high $l$, that the contributions to $F^\self_\alpha$ will drop off as $l^{-2}$. It is possible to add higher-order regularization terms to the series that increase the convergence rate with $l$. These terms are known to take the form \cite{Detweiler-Messaritaki-Whiting}
\begin{equation}\label{eq:D_alpha2n}
	\frac{D_{\alpha,2}}{(2l-1)(2l+3)} + \frac{D_{\alpha,4}}{(2l-3)(2l-1)(2l+3)(2l+5)} + \dots = \sum_{n=1}^\infty D_{\alpha,2n}\left[\prod_{k=1}^n(2L-2k)(2L+2k)\right]^{-1}\c
\end{equation}
where the $D_{\alpha,2n}$ are extra regularization parameters that serve to increase the differentiability of the regular field at the particle's location (they do not affect the value of the SF as, for instance, $\sum_{l=0}^\infty[(2l-1)(2l+3)]^{-1}=0$). With the addition of each extra parameter the convergence rate of the mode-sum increases by a factor of $l^{-2}$ (the coefficients of the odd powers of $L$ are known to be zero \cite{Detweiler-Messaritaki-Whiting}). Thus knowledge of the higher-order regularization parameters is of great use in practical calculations. In principle if all the higher-order regularization parameters are known then the convergence of the mode sum becomes exponential with $l$. In particular this implies that if a component of the field does not require regularization (i.e., all regularization parameters are known to be zero) then the sum over $l$ will converge exponentially . 

\subsection{Mode-sum in Kerr spacetime}\label{sec:mode-sum:kerr}

The regularization parameters in Kerr spacetime for the scalar, electromagnetic and gravitational self-force were first derived by Barack and Ori \cite{Barack-Ori} (see Ref.~\cite{Barack-review} for an explicit derivation). The form of the regularization parameters in Kerr spacetime is rather unwieldy so I will not repeat them here. More recently Heffernan \etal~have also derived some of the higher-order regularization parameters for a particle moving along generic geodesics in Schwarzschild spacetime \cite{Heffernan-Ottewill-Wardell:Schwarz} and equatorial geodesics in Kerr spacetime \cite{Heffernan-Ottewill-Wardell:Kerr}.

There is some subtlety to implementing the mode-sum scheme in Kerr spacetime as I now discuss.  Recall from Sec.~\ref{sec:scalar_field_decomp} that in Kerr spacetime the scalar field naturally decomposes into \textit{spheroidal-harmonic} modes. The mode-sum scheme on the other hand requires \textit{spherical-harmonic} modes as input, even in Kerr spacetime\footnote{At least within its current formulation the standard mode-sum scheme requires spherical-harmonic modes as input. It may be possible to re-formulate it and regularize directly the spheroidal-harmonic modes but this has not yet been attempted. The Discussion section of Ref.~\cite{Barack-Ori-Sago} gives an overview of the difficulties involved with this approach.}. Hence, in order to regularize using the standard mode-sum approach one must first project the spheroidal-harmonic modes onto a basis of spherical harmonics. This is achieved by expanding each spheroidal harmonic in a series of spherical harmonics:
\begin{eqnarray}\label{eq:spheroidal-decomp}
	S_{\ell m}(\theta; \sigma^2) e^{im\varphi} = \sum_{l=|m|}^{\infty} b^\ell_{lm}(\sigma^2) Y_{lm}(\theta,\varphi)		\c
\end{eqnarray}
where the $\sigma$-dependent coefficients $b^\ell_{lm}$ are determined from a recursion relation found by substituting the series expansion into the angular differential equation \eqref{eq:angulareq} (see Paper I or Ref.~\cite{Hughes} for details). As Eq.~\eqref{eq:spheroidal-decomp} is a spectral expansion of a smooth function it is expected that it will converge exponentially for all values for $\sigma^2$ -- see, e.g., Ref.~\cite{Cook-Zalutskiy} for numerical examples of the rate of convergence of this series. When $\sigma^2 = 0$ the spheroidal harmonics reduce to the standard spherical-harmonics and the coefficients $b_{lm}^\ell$ reduce to the Kronecker delta $\delta_l^\ell$. Using the $b^\ell_{lm}$'s the spherical-harmonic $l$-mode contribution to the retarded force can be written as 
\begin{eqnarray}\label{eq:SSF_full}
    F_\alpha^{(\ret)l}(x) = q\nabla_\alpha \sum_{m=-l}^l \phi_{l m}(t,r) Y_{l m}(\theta,\varphi)/r   \label{eq:field-spherical-mode}   \c
\end{eqnarray}
where $\alpha=\{t,r,\varphi\}$ (I discuss the case for $\alpha=\theta$ below) and $\phi_{lm}$ is given by
\begin{eqnarray}\label{eq:phi_lm}
	\phi_{lm}(t,r) = \sum_{k=-\infty}^\infty \sum_{\ell=|m|}^\infty b_{lmk}^\ell \psi_{\ell mk}(r) e^{-i \omega_{mk} t}		\p
\end{eqnarray}
where I have defined $b_{lmk}^\ell \equiv b_{lm}^\ell(-a^2\omega^2_{mk})$. In deriving Eqs.~\eqref{eq:field-spherical-mode} and \eqref{eq:phi_lm} I have swapped the order of the infinite sums over $\ell$ and $l$. For $x\neq x_p$ this is permissible as each sum is a spectral expansion that is uniformly convergent to a finite result. In order to compute the self-force I use Eq.~\eqref{eq:ret-mode-sum} to take the limit $x\rightarrow x_p$ in which case the sum over $l$ takes a finite value at the particle.

Formally when constructing $\phi_{lm}$ one has to sum over all spheroidal $\ell$ modes. In practice this is not necessary, as the coupling between the spheroidal and spherical-harmonic modes is relatively weak for the spheroidicities encountered in this work. In Paper I it was numerically demonstrated that the contribution from a given spheroidal-harmonic $\ell m$-mode to the spherical-harmonic $lm$-modes of the field is strongly peaked around $l=\ell$ and that its contribution to other spherical-harmonic modes decreases exponentially as one moves away from this value (see Fig.~1 in Paper I). As is expected, the coupling strengthens as the magnitude of the spheroidicity, $\sigma^2$, increases.

Equation \eqref{eq:field-spherical-mode} cannot be used in its given form to compute the $F_\theta$ component of the SSF. Recall that the regularization formula \eqref{eq:ret-mode-sum} requires the full SF per spherical-harmonic $l$-mode, summed over $m$, as input. Consequently, before regularization the $Y_{lm,\theta}$ term must first be expanded onto a basis of spherical harmonics as I now discuss.

\begin{figure}
	\centering
	\includegraphics[width=8.5cm]{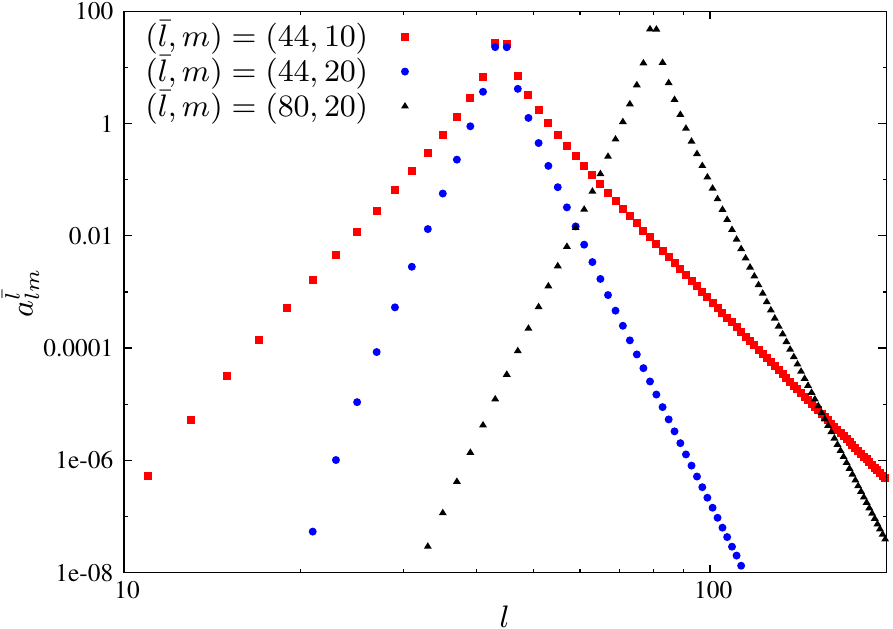}
	\caption{Projection of $Y_{\bar{l}m,\theta}$ on to a basis of spherical harmonics [see Eq.\ \eqref{eq:Y_lm_deriv_expansion}]. The contribution to each $l$-mode exhibits power-law behavior (note the $\log$-$\log$ scale). The wide bandwidth of the coupling makes practical calculations of $F_\theta$ via this method infeasible and instead I make use of the alternative approach outlined in Sec.~\ref{sec:mode-sum:kerr}.}\label{fig:Y_lm_deriv_coupling}
\end{figure}

The most na\"{i}ve route to computing $F_\theta$ is to expand $Y_{lm,\theta}$ as a series of $Y_{lm}$'s much as was done with the spheroidal harmonics [see Eq.~\eqref{eq:spheroidal-decomp}]. In this approach one would write
\begin{equation}\label{eq:Y_lm_deriv_expansion}
	Y_{\bar{l}m,\theta}(\theta,\varphi) = \sum_{l=0}^\infty a_{lm}^{\bar{l}} Y_{lm}(\theta,\varphi)	\c
\end{equation}
and then use the orthogonality of the spherical harmonics to compute the series coefficients $a_{lm}^{\bar{l}}$ via
\begin{equation}
	a_{lm}^{\bar{l}} = \oint Y_{\bar{l}m,\theta}Y^*_{lm}\, d\Omega		\c
\end{equation}
Unfortunately with this method one finds that the bandwidth of the coupling is extremely wide so that, for instance, the decomposition of $Y_{\bar{l}=44,m=10,\theta}$ couples strongly to, say, the monopole $l=0$ mode (see Fig.~\ref{fig:Y_lm_deriv_coupling}). Conversely, this means if one wishes to compute only $F_\theta^{l=0}$ at least $\bar{l}=44$ modes must be calculated. This makes any numerical computation impractical with this method. A more efficient technique for computing $F_\theta$ is to multiply the scalar field by a suitable function $f(\theta)$ that has the properties: (i) when taking the derivative with respect to $\theta$ and then the limit $\theta\rightarrow\theta_p$ the correct self-force is recovered, and (ii) the combination $f(\theta)Y_{lm,\theta}$ can be expanded in a finite series of spherical harmonics. After some experimentation, one such function that presents itself is
\begin{equation}
	f(\theta) = \frac{3\sin^2\theta_p \sin\theta - \sin^3\theta}{2\sin^3\theta_p} = 1+\mathcal{O}(\theta-\theta_p)^2		\p
\end{equation}
This function satisfies condition (i), as $(fY_{lm})_{,\theta} = fY_{lm,\theta}+f_{,\theta}Y_{lm} \rightarrow Y_{lm,\theta}$ as $\theta\rightarrow\theta_p$. Furthermore, using the identities \eqref{eq:spherical_Y_lm_ident3} and \eqref{eq:spherical_Y_lm_ident6} $fY_{lm,\theta}$ can be expanded as a series that couples only to the $l\pm1$ and $l\pm3$ modes. Performing the expansions gives the final result:
\begin{equation}\label{eq:F_theta_full}
    F_\theta^{(\ret)l}(x\rightarrow x_p) = q \sum_{m=-l}^l \phi_{l m}(t_p,r_p) \mathcal{F}_{lm}(\theta_p) Y_{l m}(\theta_p,\varphi_p)/r      \c
\end{equation}
where $\phi_{lm}$ is given by Eq.~\eqref{eq:phi_lm} and $\mathcal{F}_{lm}$ takes the form
\begin{equation}
	\mathcal{F}_{lm}(\theta_p) = \frac{3}{2\sin\theta_p}\left( \delta^{l-1,m}_{(+1)} + \delta^{l+1,m}_{(-1)}\right) - \frac{1}{2\sin^3\theta_p} \left(\zeta^{l-3,m}_{(+3)} + \zeta^{l-1,m}_{(+1)} + \zeta^{l+1,m}_{(-1)} + \zeta^{l+3,m}_{(-3)} \right)		\c
\end{equation}
with the $\delta$'s and $\zeta$'s given in Appendix \ref{apdx:identities}.

It is worth noting that for electromagnetic and gravitational self-force calculations the orthogonality of the self-force and the four-velocity ($u^\alpha F^{\text{EM/Grav}}_\alpha = 0$) can be employed, from which, once $F_t, F_\varphi$ and $F_r$ are known, one can compute $F_\theta$. However recall from Sec.~\ref{sec:equations_of_motion} that for the scalar field setup used in this work the quantity $u^\alpha F_\alpha$ is generally non-zero and thus $F_\theta$ must be computed directly using a method such as the one given above.

\subsection{Conservative and dissipative self-forces}

For bound geodesic motion in Kerr spacetime the self-force can be uniquely separated into conservative and dissipative components\footnote{For an evolving, non-geodesic orbit the conservative/dissipative split is not well defined \cite{Diener_etal:SSF_self_consistent}}. In this work it will suffice to give the formulae for their construction --- see Papers I and II for a more in depth discussion. For a review of the effects of these two components on an inspiral see Refs.~\cite{Pound-Poisson:Osculating_orbits,Warburton_etal}

Taking $\chi=0$ to be the turning point of the polar motion, the $l$-mode contributions to the (retarded) conservative and dissipative components of the self-force can be constructed via \cite{Barack-review}
\begin{align}\label{eq:Frl_cons_diss}
	F^{l(\cons)}_\alpha(\chi) = \frac{1}{2}\left[ F^{l(\ret)}_\alpha(\chi) + \epsilon_{(\alpha)} F^{l(\ret)}_\alpha(-\chi)\right]\,,\qquad F^{l(\diss)}_\alpha(\chi) = \frac{1}{2}\left[ F^{l(\ret)}_\alpha(\chi) - \epsilon_{(\alpha)} F^{l(\ret)}_\alpha(-\chi)\right]\c
\end{align}
where $\epsilon_{(\alpha)} = (-1,1,1,-1)$ in Boyer-Lindquist coordinates. The regularized conservative and dissipative self-forces are then constructed with
\begin{align}\label{eq:Frl_cons_diss_reg}
	F^{\cons}_\alpha = \sum_{l=0}^\infty \left( F^{l(\cons)}_{\alpha \pm} - A_{\alpha\pm} L - B_\alpha - C_\alpha L^{-1} \right)\equiv \sum_{l=0}^\infty F^{l(\cons,R)}_\alpha  \c\qquad F^{\diss}_\alpha = \sum_{l=0}^\infty F^{l(\diss)}_\alpha\p
\end{align}
As discussed in Sec.~\ref{sec:SF_via_mode_sum}, because the dissipative self-force does not require regularization the sum over $l$ converges exponentially. The conservative component on the other hand converges as $l^{-2}$ unless higher-order regularization parameters are employed (the extra parameters are the same as for $F^{(\ret)l}_\alpha$). Owing to the different convergence rates, splitting the self-force into conservative and dissipative quantities is practically beneficial when it comes to estimating the contribution from the uncomputed $l$ modes -- see Paper II for a discussion.

\section{Numerical implementation}\label{sec:numerical_implementation}

In the following section I give an overview of the steps required in the calculation of the SSF for inclined circular orbits. The calculation presented here builds on that of Papers I and II and so, where appropriate, I will refer to those works for the sake of brevity. In particular the construction of the appropriate boundary conditions and resulting homogeneous fields is essentially identical between the three articles.

\subsection{Numerical Boundary conditions}\label{sec:numerical_BCs}

Let the numerical domain extend from $r_*=r_{*\in} \ll -M$ to $r_*=r_{*\out} \gg M$. The asymptotic form of the (retarded-field) boundary conditions at spatial infinity and the event horizon are discussed in Sec.~II.~C.~of Paper I. The numerical boundary conditions are then constructed by expanding the asymptotic boundary conditions in the following manner
\begin{align}
	\psi^+_{\ell mk}(r_\out) 	&= e^{+i\omega_{mk} r_{*\out}} \sum_{n=0}^{\bar{n}_\out} c_n^+ r_\out^{-n}\c		\label{eq:outer_boundary_expansion}\\
	\psi^-_{\ell mk}(r_\in) 	&= e^{-i\gamma_{mk} r_{*\in}} \sum_{n=0}^{\bar{n}_\in} c_n^- (r_\in - r_+)^n\c		\label{eq:inner_boundary_expansion}
\end{align}
with $r_+=M+\sqrt{M^2-a^2}$ as the location of the event horizon, $\gamma_{mk} = (2Mr_+ \omega_{mk} - am)/r_+^2$, $r_\in=r(r_{*\in})$ and $r_\out=r(r_{*\out})$. The series coefficients $c_n^\pm$ are determined by substituting the above forms for $\psi^\pm_{\ell mk}$ in to Eq.~\eqref{eq:fieldeq}. The resulting recursion relations are rather lengthy --- see Appendix C of Paper I for their explicit form. As the mode frequency, $\omega_{mk}$, differs for each $\ell mk$-mode the extension of the numerical domain depends on the particular $\ell mk$-mode under consideration. I find that setting $r_{*\in}=-50M$ is, for all modes, sufficient to ensure rapid convergence of the expansion near the inner boundary. Though I have not been explicit in Eq.~\eqref{eq:outer_boundary_expansion}, the outer boundary is formally an expansion in $(\omega_{mk} r_\out)^{-1}$. Consequently, one must arrange for $\omega_{mk}r_\out  \gg 1$ to ensure rapid convergence of the outer boundary condition. Lastly, I choose the truncation parameters $\bar{n}_{\in/\out}$ such that the contribution from the next term in the series drops below a relative threshold of $10^{-12}$ compared to the first term in the series (which I take to be $c^\pm_0=1$).

\subsection{Junction conditions}

The standard variation of parameters approach can be used to construct the inhomogeneous solutions from the homogeneous fields. Let $\tilde{\psi}^{\pm}_{\ell mk}$ denote the homogeneous solutions to the radial equation \eqref{eq:radialeq} obtained by setting $c_0^\pm=1$ in the boundary conditions \eqref{eq:outer_boundary_expansion} and \eqref{eq:inner_boundary_expansion}. The inhomogeneous radial solutions, $\psi^{\pm}_{\ell m}$, are then constructed via
\begin{align}
	\psi^\pm_{\ell mk}(r) = \alpha_{\ell mk} \frac{\tilde{\psi}_{\ell mk}^{\mp}(r_0)}{\tilde{W}_0} \psi^\pm_{\ell mk}(r)\c		\label{eq:inhom_from_hom}
\end{align}
where $\tilde{W}_0 = \tilde{\psi}^-_{\ell mk}(r_0)\tilde{\psi}_{\ell mk}^{+\prime}(r_0) - \tilde{\psi}^+_{\ell mk}(r_0)\tilde{\psi}_{\ell mk}^{-\prime}(r_0) $ is the Wronskian of the homogeneous solutions with a prime denoting differentiation with respect to $r_*$. The coefficient, $\alpha_{\ell m}$, in Eq.~\eqref{eq:inhom_from_hom} represents the jump in the $r_*$ derivative across $r=r_0$ and is given explicitly by
\begin{align}
	\alpha_{\ell mk} = (\psi^{+\prime}_{\ell mk} - \left.\psi
^{-\prime}_{\ell mk})\right|_{r_0} = -\frac{4\pi q}{r_0 T_\theta}\int^{2\pi}_0 \frac{S_{\ell m}(\theta_p(\chi);-a^2\omega_{mk}^2)}{u^t(\theta_p(\chi))} \cos(\omega_{mk} t_p(\chi) - m\varphi_p(\chi))\frac{dt}{d\chi}\,d\chi\p
\end{align}

\subsection{Algorithm}

In this section I outline the explicit steps in my numerical calculation

\begin{itemize}
	\item{\textit{Orbital parameters.} For a given black hole spin, $a$, orbital radius, $r_0$, and inclination angle, $\iota$, calculate the quantities related to the associated inclined circular geodesic ($\en,\ang,\Carter,\Omega_\varphi,\Omega_\theta, T_\theta$, etc.) using the formulae given in Sec.~\ref{sec:orbit_param}}.

	\item{\textit{Numerically solve the radial field equation.} For each radiative $(\omega \neq 0)$ $\ell mk$-mode construct boundary conditions at the edges of the numerical domain using the procedure outlined in Sec.~\ref{sec:numerical_BCs}.  Next integrate from the boundaries to $r=r_0$ using the integration routines outlined in Paper I. For the static ($\omega=0$) modes the radial equation \eqref{eq:radialeq} admits analytic solutions which are given explicitly in Paper I. For each $\ell mk$-mode construct the inhomogeneous solutions using Eq.~\eqref{eq:inhom_from_hom} and store the value of $\psi_{\ell mk }$ and its (one-sided) $r$ derivatives along the orbit.}

	\item{\textit{Spheroidal to spherical-harmonic decomposition.} Using the values of the field and its derivatives for each $\ell mk$-mode (up to some maximum $\ell=\ell_\max$) construct the $\alpha=r,t,\varphi$ spherical-harmonic $l$-mode contributions to the full retarded force, $F_\alpha^{(\ret)l}$, using Eq.~\eqref{eq:field-spherical-mode}. For $F_\theta^{(\ret)l}$ Eq.~\eqref{eq:F_theta_full} should be used. Let the greatest spherical-harmonic $l$-mode which does not have a contribution of more than $10^{-12}$ (relatively) from an uncomputed spheroidal-harmonic $\ell$-mode be denoted by $l_\max$.  For the orbits encountered in this work the coupling between the spheroidal and spherical-harmonic modes is not particularly strong and typically $|\ell_{\max} - l_{\max}| < 10$. In general for orbits with larger radii and inclinations nearer to the equatorial plane the coupling is weaker. The coupling is also weaker for smaller values of $a$.  }

	\item{\textit{Determine $k_{\max}$.} Formally, for each $\ell m$-mode, one must sum over all $k$-modes. In practice I find for large $k$ the contribution from each $k$-mode drops off exponentially. Consequently I truncate the sum over $k$ once the contribution to the scalar-field and its $t$-, $r$-, $\theta$- and $\varphi$-derivatives drops below $10^{-12}$. In general $k_{\max}$ is greater for modes with higher spheroidicity ($=-a^2\omega^2$). As an example, consider the orbit with parameters $(a,r_0,\ang) = (0.998,4,0.5)M$ which is the strongest-field, highest inclination ($\iota\approx81.03^\circ$) orbit I consider in this work. In this case the greatest $k_{\max}$ encountered is for the $\ell=40,m=35$ mode with $k_{\max}=79$. In general, orbits with lower inclinations and/or larger orbital radii have lower values of $k_{\max}$.}

	\item{\textit{Regularization of the spherical-harmonic $l$-modes.} Compute the conservative and dissipative self-forces via Eq.~\eqref{eq:Frl_cons_diss_reg}. The dissipative component of the self-force does not require regularization and I find that summing the first 15 $l$-modes is sufficient to compute the dissipative sector with a relative accuracy of $10^{-9}$. For the conservative component the slow, power-law convergence of the sum with $l$ necessitates extrapolating the contribution from the uncomputed $l$-modes. I estimate this contribution using the method detailed in Paper I. From Eq.~\eqref{eq:Frl_cons_diss}, the total self-force is simply the sum of the conservative and dissipative pieces.  }
\end{itemize}

My code is parallelized to run on a cluster using the Message Passing Interface (MPI). Each processing core computes the necessary $k$-modes for a given $\ell m$-mode that is dynamically assigned to it. Once all the $\ell m$-modes with $\ell<\ell_\max$ are computed the results from each core are combined to give the final result.

\section{Results}\label{sec:results}

\begin{figure}[htb]
	\centering
    \includegraphics[width=8.0cm]{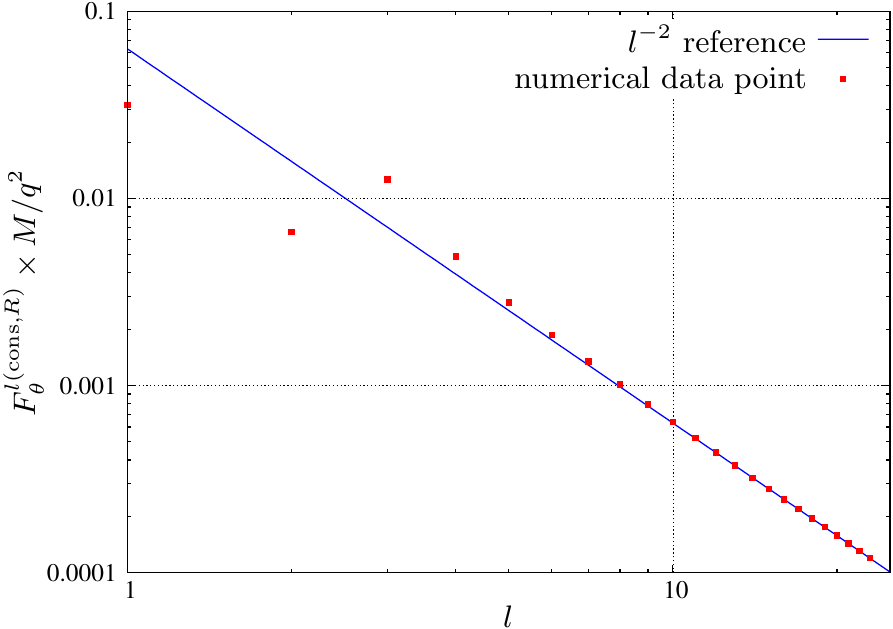}\hspace{1cm}
    \includegraphics[width=8.0cm]{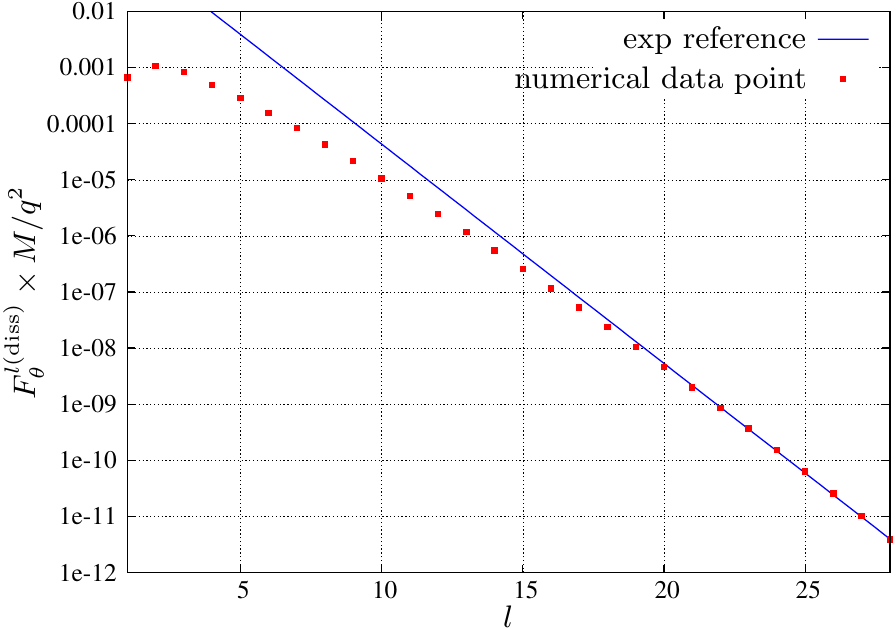}
    \caption[Convergence of $F_\theta^{l(\cons/\diss)}$ components of the SSF for a sample inclined circular orbit]{Convergence of $F_\theta^{l(\cons/\diss)}$ components of the SSF for a sample inclined circular orbit about a black hole with $a=0.998M$ and with orbital parameters $(r_0,\iota)=(3M, 27.5573^\circ)$, shown at $\chi=0.816814$. The left panel depicts the contributions per $l$-mode to the conservative component of $F_\theta$ alongside an $l^{-2}$ reference line. As theory predicts, the contribution of the high $l$-modes follows closely to the reference line. The right panel shows the contributions per $l$-mode to the dissipative component of $F_\theta$ alongside an exponential reference line. Again, as expected from theory, the contribution from the high $l$-modes follows closely to this line. Similar convergence behavior is observed for the conservative and dissipative pieces of the other three components ($F_r, F_t, F_\varphi$) of the SSF.}\label{fig:Ftheta_l_convergence}
\end{figure}

Before considering results for orbits in Kerr spacetime I present two validation tests that demonstrate that my code is performing as desired. First, as discussed in Sec.~\ref{sec:SF_via_mode_sum}, with the known regularization parameters the high-$l$ mode contribution to the SSF should fall off as $l^{-2}$. I observe this behavior in all my numerical data --- see Fig.~\ref{fig:Ftheta_l_convergence} for an example. As a second test on my code I compute the SSF for inclined circular orbits about a Schwarzschild black hole as I outline in the following subsection. Throughout this section I shall often use an over-tilde to denote an adimensionalized quantity, e.g., $\tilde{F}_{\{t,r\}} = (M^2/q^2)F_{\{t,r\}}$ and $\tilde{F}_{\{\theta,\varphi\}} = (M/q^2)F_{\{\theta,\varphi\}}$.

\subsection{Inclined circular orbits about a Schwarzschild black hole}

Owing to the spherical symmetry of the spacetime, geodesic orbits about a Schwarzschild black hole can, without loss of generality, be arranged to lie within the equatorial plane. Making this choice often simplifies the resulting calculation. The components of the SSF for an orbit out of the equatorial plane can be simply related, via a rotation of the coordinate system, to the components of the SSF for an orbit in the equatorial plane. These relations are given by
\begin{align}
	F_t(\iota)					&= F_t^\text{eq}\c													\label{eq:F_t_Schwarz}					\\
	F_r(\iota)					&= F_r^\text{eq}\c													\\
	F_\varphi(\iota) 			&= (\cos\iota) F_\varphi^\text{eq}\c								\\
	F_\theta(\iota,\theta_p) 	&= \pm F_\varphi(\iota) \sqrt{\sec^2\iota - \csc^2\theta_p}\c		\label{eq:F_phi_Schwarz}
\end{align}
where the `eq' superscript denotes a quantities value for an equatorial orbit ($\iota = 0$) and the $\pm$ has the same meaning as in Eq.~\eqref{eq:dtheta_dt}. The $t$-, $r$- and $\varphi$-components of the SSF are constants along the inclined orbit. The first two take the same value as in the equatorial case, the third is related by a multiplicative constant for each inclination. The $\theta$-component of the SSF varies along the inclined orbit.  These equations can be used to test the numerical results in the $a=0$ case. Table \ref{tbl:Schwarz} gives some sample results showing the above equations hold to a relative accuracy of $\lesssim 10^{-7}$.

When making the calculation in Schwarzschild spacetime recall that orbital parametrization for orbits about a Kerr black hole presented in Sec.~\ref{sec:orbit_param} breaks down and the equations in Sec.~\ref{sec:orbit_param_Schwarz} must be used instead. It is also worth noting that, as $\Omega_\theta=\Omega_\varphi$ for inclined circular orbits in Schwarzschild spacetime, the $k=-m$ modes are static as $\omega_{mk}=m\Omega_\varphi + k\Omega_\theta=0$ in this case. Consequently, for these modes the static solutions given in Paper I can be used.

\begin{table}
\begin{tabular}{l|ll|l}
\toprule
			&	$ \tilde{F}_\alpha(\iota\simeq38.86^\circ)$ &	$\tilde{F}_\alpha^\text{eq}$			&	 $|\text{rel.~error*}|$				\\
\hline
$\tilde{F}_t$		&	$8.77446723272\times10^{-5}$	&	$8.77446723265\times10^{-5}$		&	$7.8\times10^{-12}$			\\
$\tilde{F}_r$		&	$3.6188115\times10^{-5}$		&	$3.6188106\times10^{-5}$			&	$2.5\times10^{-7}$			\\
$\tilde{F}_\theta$	&	$-1.29276938\times10^{-3}$		&	$0$									&	$4.6\times10^{-8}$			\\
$\tilde{F}_\varphi$	&	$-1.604306657\times10^{-3}$		&	$-2.060352528\times10^{-3}$			&	$2.98\times10^{-8}$			\\
\hline

\botrule
\end{tabular}
\caption{The adimensionalized SSF for inclined and equatorial circular orbits with radius $r_0=8.2M$ about a Schwarzschild black hole. The values for $\tilde{F}_\theta$ are given at $\chi=\pi/2$. The second column shows the SSF for an inclined orbit with $\ang=2.8M$ $(\iota\simeq38.86^\circ)$. The third column shows the SSF for an equatorial orbit ($\iota=0^\circ$) computed using the code presented in Paper I. The forth column shows relative difference between the left- and right-hand sides of Eqs.~\eqref{eq:F_t_Schwarz}-\eqref{eq:F_phi_Schwarz} computed using the data in columns two and three. The results presented in this table were computed with $\ell_{\max}=l_{\max}=40$. }
\label{tbl:Schwarz}
\end{table}

\subsection{Inclined circular orbits about a Kerr black hole}

\begin{figure}
	\includegraphics[width=10cm]{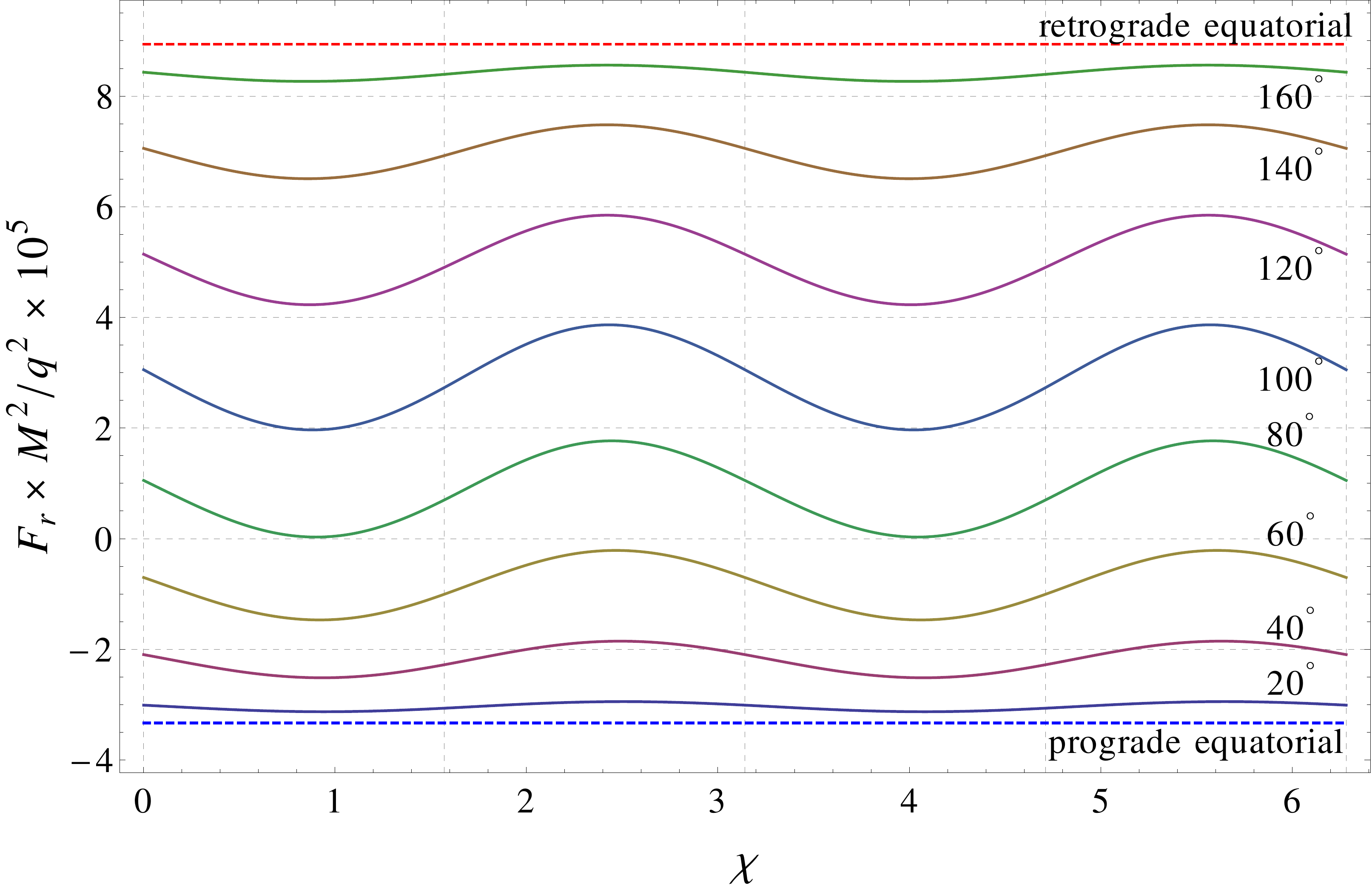}
	\caption{The radial SSF for various inclined circular orbits with radius $r_0=9M$ in motion about a black hole with spin parameter $a=0.998M$. The self-force varies smoothly from the prograde equatorial orbit ($\iota=0^\circ$) to the retrograde equatorial orbit ($\iota=180^\circ$) with the largest oscillations observed for near-polar orbits. The equatorial prograde and retrograde values were computed using the code presented in Paper I.}\label{fig:Fr_r9M}
\end{figure}

\begin{figure}[htb]
	\centering
    \includegraphics[width=11.0cm]{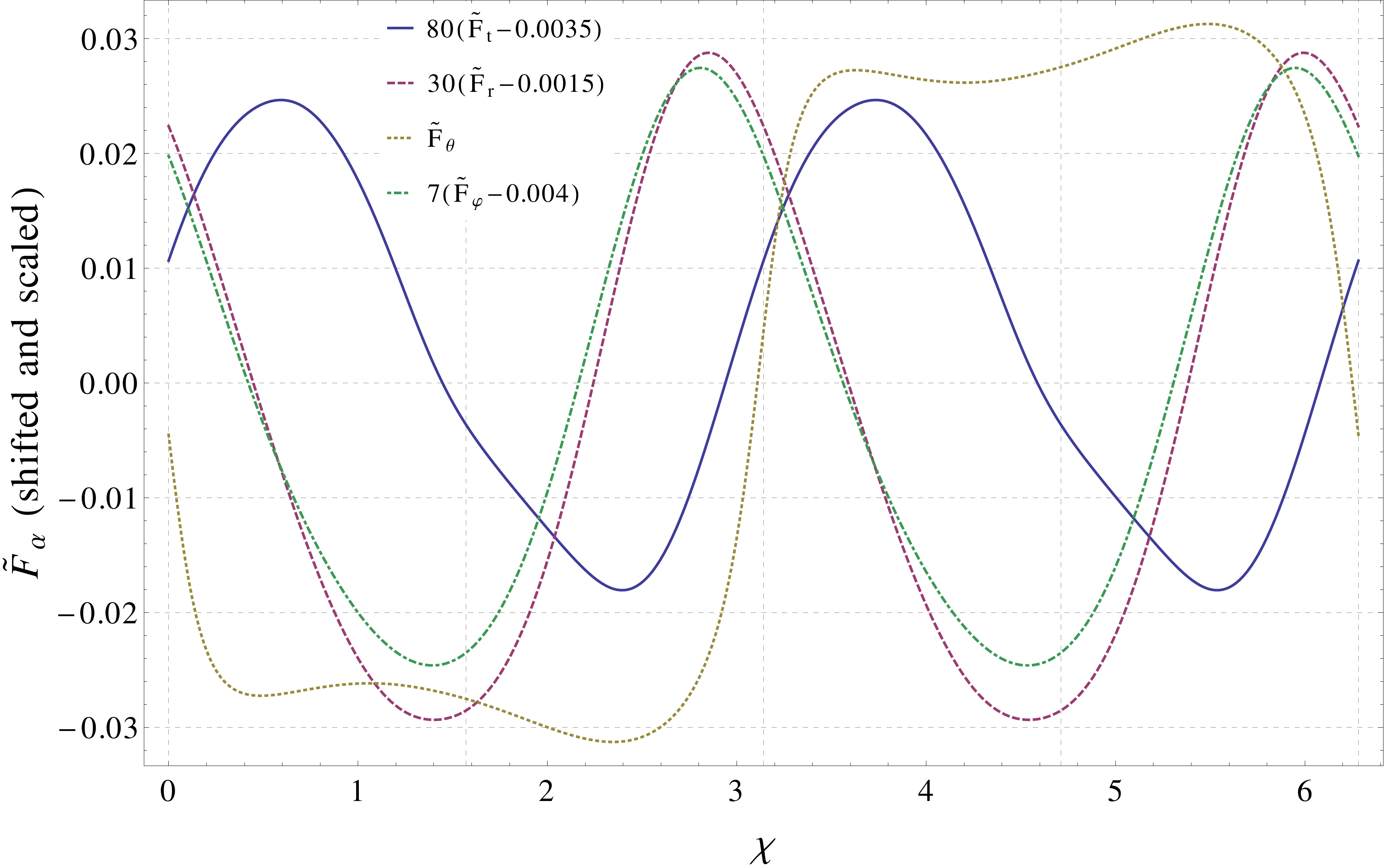}
	\caption{The SSF for an inclined circular orbit of radius $r_0=4M$ about a Kerr black hole with spin $a=0.998M$. The angular-momentum of the orbit is $\ang=-0.9M$ $(\iota\simeq102.81)$. To make clear the relative phasing between the different components of the SSF $\tilde{F}_t,\tilde{F}_r$ and $\tilde{F}_\varphi$ have been shifted and rescaled as shown in the legend.}\label{fig:SSF_circ_inclined}
\end{figure}

\begin{figure}
	\includegraphics[width=8cm]{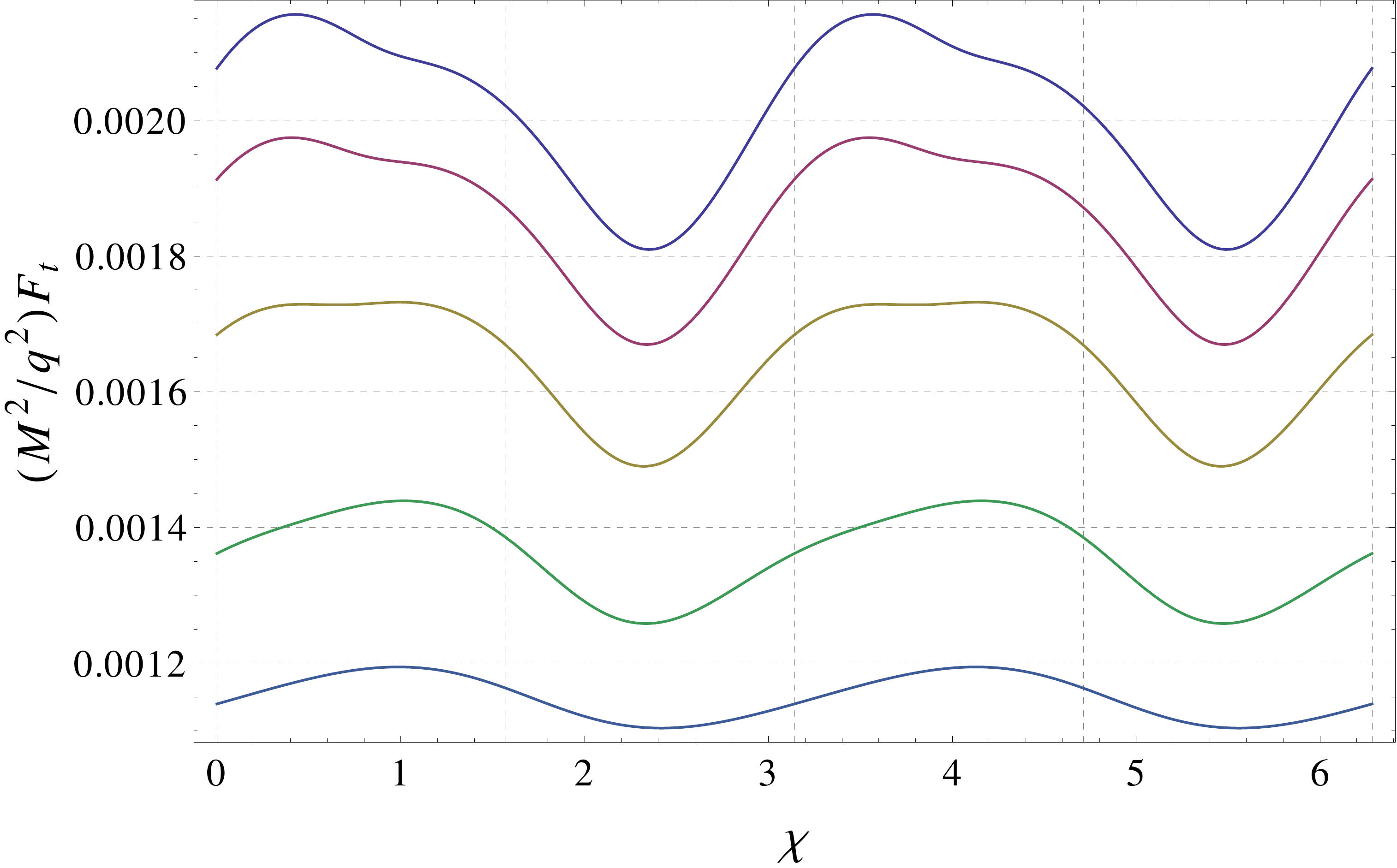}\hspace{0.5cm}
	\includegraphics[width=8cm]{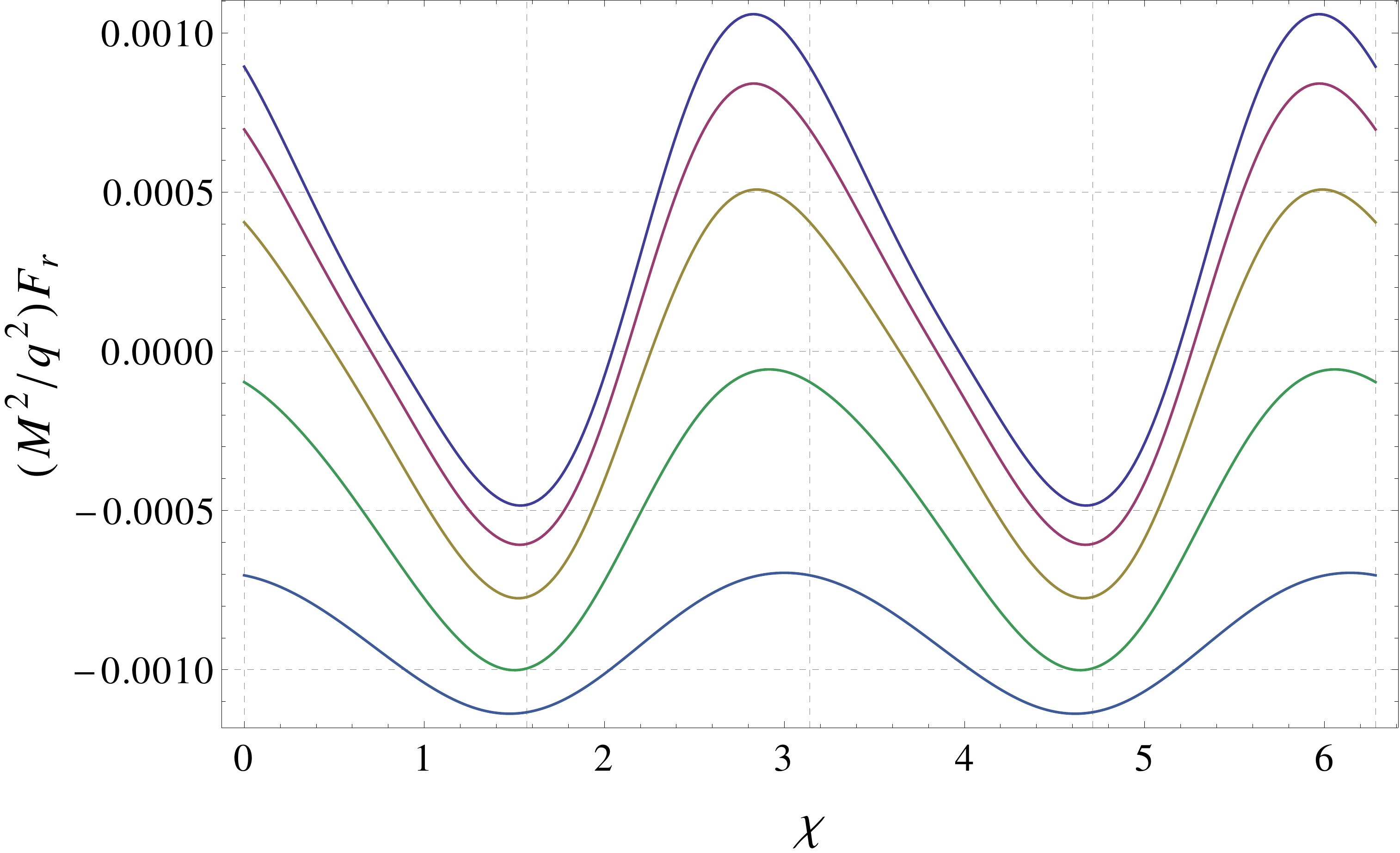}
	\includegraphics[width=8cm]{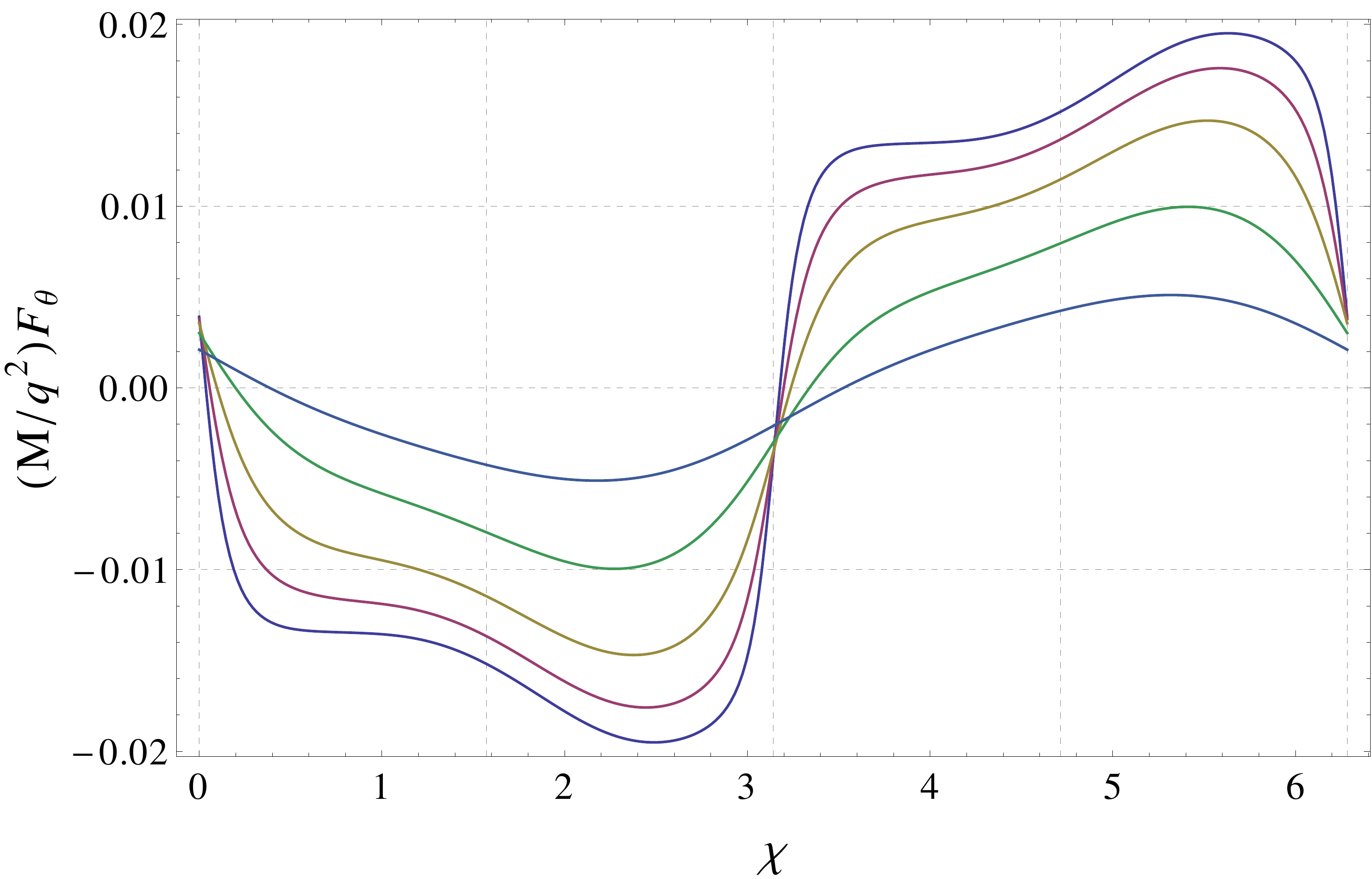}\hspace{0.5cm}
	\includegraphics[width=8cm]{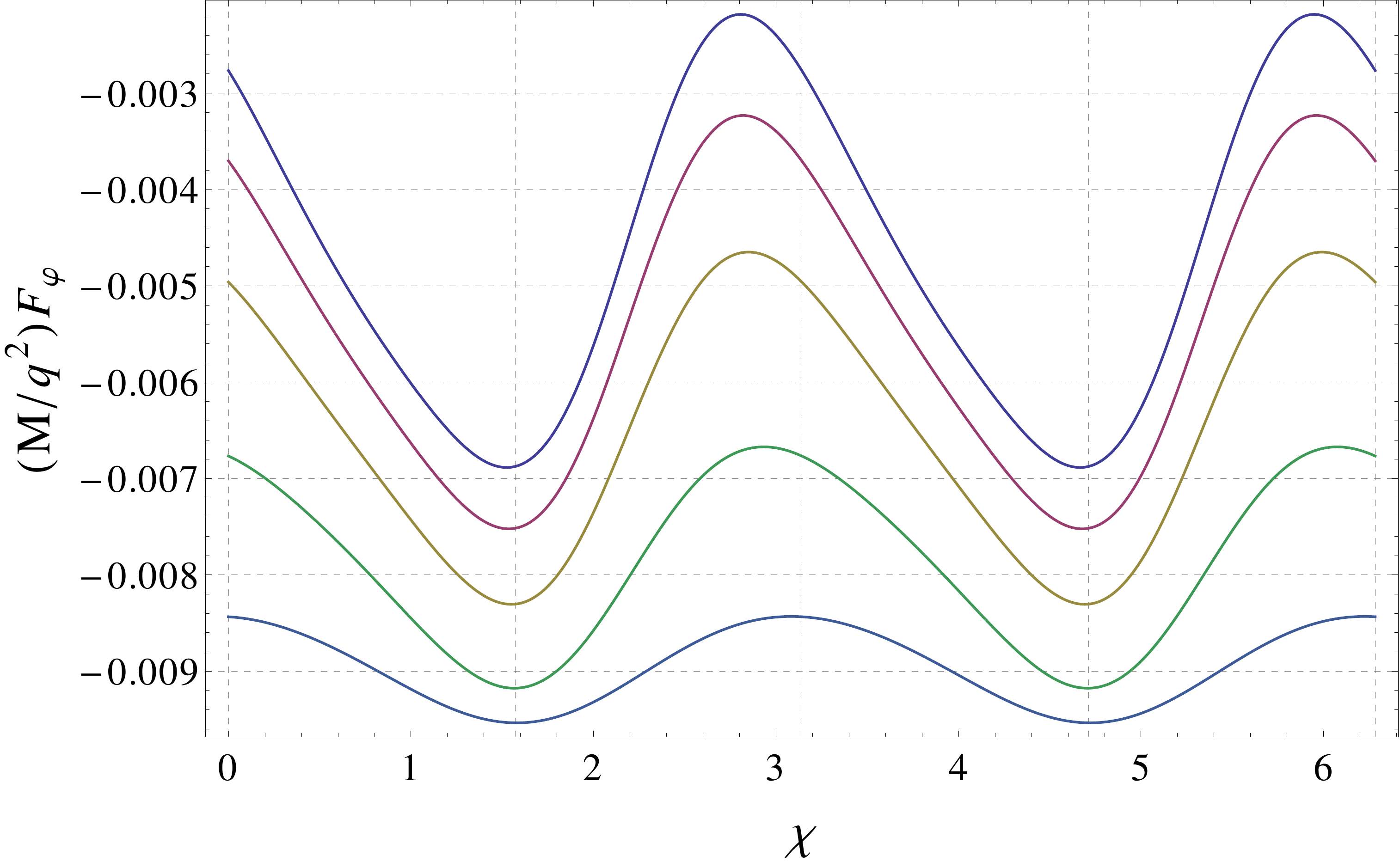}
	\caption{The four components of the covariant SSF for a black hole with spin $a=0.998M$ and orbit of radius $r_0=4M$. Reading clockwise from the top-left the figures show the adimensionalized $F_t, F_r, F_\varphi$ and $F_\theta$. In each plot the different curves correspond to orbits with angular-momentum of $\ang=\{0.5, 0.7, 1, 1.5, 2\}$, reading from top-to-bottom when $\chi>\pi$. These correspond to orbits with inclinations of $\iota\approx 81.03\degree, 76.93\degree, 70.06\degree, 55.87\degree, 34.73\degree$ respectively. }\label{fig:a0.998_r4}
\end{figure}

In this section I present some sample results for the SSF experienced by a particle moving on an inclined circular orbit about a Kerr black hole. Explicit numerical values can be found in Tables \ref{tbl:data_diss} and \ref{tbl:data_cons}. The force along a variety of orbits is plotted in Figures \ref{fig:Fr_r9M}-\ref{fig:a0.998_r4}. For the highest-spin, strongest-field, nearest-polar orbit considered in this work [$(a,\iota,r_0)=(0.998M, 81.03, 4)$] the computation takes approximately 12 hours on 12 cores of a 3GHz cluster. Results for lower spins, near-equatorial inclinations and larger radii orbits take less time. My current code is unable to compute the SSF for precisely polar obits ($\iota=90^\circ$) as the orbit parameterization I have used breaks down there.

I observe that the $t$-, $r$- and $\varphi$-components of the SSF have a period equal to half the orbital period. The period of the $\theta$-component is the same as the orbital period. This is expected from geometrical considerations which give $(F_r, F_t, F_\theta,F_\varphi)\rightarrow(F_r,F_t,-F_\theta,F_\varphi)$ as $\chi\rightarrow\chi+\pi$. This in turn implies the observed periodicity of the components of the SSF. Figure \ref{fig:SSF_circ_inclined} shows the phasing between the various components for a sample orbit. For all the orbits I have examined I observe that the $r$- and $\varphi$-components of the SSF are roughly in phase whereas the $t$-component is not similarly synchronized (see, e.g., Fig.~\ref{fig:SSF_circ_inclined}). In general, unlike for the case of inclined orbit about a Schwarzschild black hole, the $\theta$-component is non-zero at $\chi=0,\pi,2\pi,..$ (i.e., at $\theta_p=\theta_{\min/\max}$).

\begin{table}
\begin{tabular}{lll|llll}
\toprule
$\ang/M$ 	& $\iota$		& $\chi$ 	& $\tilde{F}^\diss_t\times10^{3}$		& $\tilde{F}^\diss_r\times10^{4}$		& $\tilde{F}^\diss_\theta\times10^{2}$	& $\tilde{F}^\diss_\varphi\times10^{3}$	\\
\hline
$-0.9$			& $102.81^\circ$	& $0$		& $3.6331386$ 	& $0$				& $0$				& $6.8251604$	\\
				&					& $\pi/3$	& $3.5097700$ 	& $-2.5305833$ 		& $-2.83402790$		& $2.1971529$ 	\\
				&					& $\pi/2$	& $3.4544807$	& $0$				& $-2.75171580$		& $0.6429822$ 	\\
\hline
$0.5$			& $81.03^\circ$		& $0$		& $2.0765973$	& $0$				& $0$				& $-2.7658579$	\\
				&					& $\pi/3$	& $1.9708447$	& $-1.4903886$		& $-1.5973433$		& $-5.6107744$	\\
				&					& $\pi/2$	& $2.0210348$	& $0$				& $-1.5184616$		& $-6.8756152$	\\
\hline
$1.0$			& $70.06^\circ$		& $0$		& $1.683771$ 	& $0$				& $0$				& $-4.9608699$	\\
				&					& $\pi/3$	& $1.623585$	& $-1.2804071$ 		& $-1.1852125$		& $-7.2462960$ 	\\
				&					& $\pi/2$	& $1.6686414$	& $0$ 				& $-1.1462029$ 		& $-8.3045156$	\\
\hline
$1.5$			& $55.87^\circ$		& $0$		& $1.3614277$ 	& $0$				& $0$				& $-6.7669429$	\\
				&					& $\pi/3$	& $1.3570188$	& $-0.93549168$ 	& $-0.78780591$ 	& $-8.4312270$	\\
				&					& $\pi/2$	& $1.3851019$	& $0$ 				& $-0.79457458$ 	& $-9.1771681$ 	\\
\hline
$2.0$			& $34.73^\circ$		& $0$		& $1.1396681$ 	& $0$				& $0$				& $-8.4356316$	\\
				&					& $\pi/3$	& $1.1541704$ 	& $-0.42965916$ 	& $-0.39033772$ 	& $-9.2293270$ 	\\
				&					& $\pi/2$	& $1.1629781$ 	& $0$ 				& $-0.42358489$ 	& $-9.5365703$ 	\\
\hline
$\ang^\text{pro}$ & $0^\circ$ 	& - 	& $1.0592881$	& $0$ 				& $0$ 				& $-9.5314743$ 	\\
\botrule
\end{tabular}
\caption{Sample results for the dissipative SSF for a strong-field orbit with $r_0=4M$ about a black hole with spin $a=0.998M$. The various components of the total SSF for these orbits are plotted in Figs.~\ref{fig:SSF_circ_inclined} and \ref{fig:a0.998_r4}. The values for $\chi>\pi$ can be inferred from Eq.~\eqref{eq:Frl_cons_diss}. The final row shows the result for the prograde circular orbit in the equatorial plane ($\ang^\text{pro}\simeq2.999551M$) calculated using the code presented in Paper I. The retrograde value is not given as there are no stable retrograde circular equatorial orbits with $r_0<9M$. The results presented in this table were computed with $\ell_{\max}=40$. All digits presented are accurate.}
\label{tbl:data_diss}
\end{table}

\begin{table}
\begin{tabular}{lll|llll}
\toprule
$\ang/M$ 	& $\iota$			& $\chi$ 	& $\tilde{F}_t^\cons\times10^{-4}$		& $\tilde{F}^\cons_r\times10^{3}$		& $\tilde{F}^\cons_\theta\times10^{-3}$	& $\tilde{F}^\cons_\varphi\times10^{-3}$	\\
\hline
$-0.9$				& $102.81^\circ$	& $0$		& $0$						& $2.24568(4)$			& $-4.5224(9)$				& $0$						\\
					&					& $\pi/3$	& $1.8996(8)$				& $0.91201(7)$			& $2.1818(9)$				& $-1.19543(6)$				\\
					&					& $\pi/2$	& $0$						& $0.54824(9)$			& $0$						& $0$						\\
\hline
$0.5$				& $81.03^\circ$		& $0$		& $0$						& $0.89359(3)$			& $3.9080(9)$				& $0$						\\
					&					& $\pi/3$	& $0.11977(2)$				& $-0.05559(5)$			& $2.3730(3)$				& $-0.52212(2)$				\\
					&					& $\pi/2$	& $0$						& $-0.48234(4)$			& $0$						& $0$						\\
\hline
$1.0$				& $70.06^\circ$		& $0$		& $0$			 			& $0.405036(4)$			& $3.55243(9)$				& $0$ 						\\
					&					& $\pi/3$	& $1.07740(5)$				& $-0.390190(8)$		& $2.25495(4)$				& $-0.29793(5)$				\\
					&					& $\pi/2$	& $0$						& $-0.772001(4)$		& $0$						& $0$						\\
\hline
$1.5$				& $55.87^\circ$		& $0$		& $0$						& $-0.096867(3)$		& $3.01413(2)$				& $0$						\\
					& 					&$\pi/3$	& $0.81807(4)$				& $-0.716746(5)$		& $1.89867(1)$				& $-0.11192(3)$				\\
					&					& $\pi/2$	& $0$						& $-0.995883(2)$		& $0$						& $0$						\\
\hline
$2.0$				& $34.73^\circ$		& $0$		& $0$						& $-0.7036066(5)$		& $2.097350(7)$				& $0$						\\
					&					& $\pi/3$	& $0.396453(2)$				& $-1.0148358(3)$		& $1.184335(2)$				& $-0.0044904(3)$			\\
					&					& $\pi/2$	& $0$						& $-1.1335073(3)$		& $0$						& $0$						\\
\hline
$\ang^\text{pro}$ & $0^\circ$ 			& - 		& $0$						& $ -1.1687088(7)$ 		& $0$ 						& $0$ 						\\
\botrule
\end{tabular}
\caption{The same as Table \ref{tbl:data_diss} but for the conservative SSF. The numbers in brackets shows the estimated error in the last digit presented, i.e., $2.1852(8)=2.1852\pm8\times10^{-4}$. The power-law convergence of the $l$-modes in the conservative sector necessitates estimating the contribution from the uncomputed $l$-modes. This, in turn, leads to the results for the conservative SSF being less accurate than those presented for the dissipative sector. The results presented in this table were computed with $\ell_{\max}=40$.}
\label{tbl:data_cons}
\end{table}

\section{Concluding remarks}

In this work I have computed the scalar-field self-force (SSF) experienced by a particle moving along an inclined circular geodesic orbit about a Kerr black hole. I validated my code by confirming that the high-$l$ contributions to the SSF fall off as theory predicts and that the code performs as expected in the Schwarzschild limit. I have presented results for the dissipative and conservative pieces of all four components of the SSF for a particle moving along strong-field geodesic orbits about a rapidly rotating black hole. This is the first time a conservative self-force has been computed for a non-equatorial orbit in Kerr spacetime. For the scalar-field studied in this work the orthogonality relation $u^\alpha F_\alpha = 0$ does not hold (see Sec.~\ref{sec:equations_of_motion}) and thus the $\theta$-component of the SSF must be computed directly. Whilst in the electromagnetic and gravitational cases one can avail of the orthogonality relation it might be helpful to calculate all the components of the self-force directly using the method presented in Sec.~\ref{sec:mode-sum:kerr} and use the orthogonality relation as a check on the accuracy of the final results.

The natural extension of this work is to consider generic bound orbits i.e., ones that are both inclined and eccentric, in Kerr spacetime. One motivation for making such an extension is that it would allow for the study of resonant orbits. These orbits occur when the the ratio of the polar and radial frequencies is a low-order rational number and are know to strongly influence the phasing of an inspiral \cite{Hinderer-Flanagan:resonances}. Schmidt has shown that a frequency-domain representation of the field equations must exist for generic orbits \cite{Schmidt} but, due to the coupled nature the radial and polar motion, it is challenging to construct such a representation in practice. Fortunately, Drasco and Hughes have overcome this difficulty by working with an alternative time coordinate (often attributed to Mino \cite{Mino}) that decouples the radial and polar motion \cite{Drasco-Hughes_FD}. Using Drasco and Hughes' technique it should be possible to compute the SSF for generic bound orbits in the frequency-domain, though the requirement to sum over three frequency indices (related to the azimuthal, polar and radial frequencies) could make the calculation for even moderately eccentric, high inclination orbits rather computationally expensive.

To conclude, it is common practice in the self-force community to develop computation techniques for the scalar-field case before attacking the gravitational problem. The high-accuracy frequency-domain results of this paper can be used as a benchmark for emerging self-force codes.  Lastly, I note that the results presented in this work for the conservative sector could be improved by employing higher-order analytic regularization parameters, but currently these are only known for geodesic motion in the equatorial plane of a Kerr black hole \cite{Heffernan-Ottewill-Wardell:Kerr}.

\section*{Acknowledgements}

I am grateful to Leor Barack and Sam Dolan for helpful discussions. I also thank Sarp Akcay and Anna Heffernan for feedback on a draft of this article. This work was supported by the Irish Research Council, which is funded under the National Development Plan for Ireland.

\appendix

\section{Spherical-harmonic identities}\label{apdx:identities}
In Sec.~\ref{sec:mode-sum:kerr} the following identities are useful \cite{Barack-Sago-eccentric}
\begin{eqnarray}
	\sin\theta Y^{lm}_{,\theta} &=& \delta^{lm}_{(+1)}Y^{l+1,m} + \delta^{lm}_{(-1)} Y^{l-1,m}		\c					\label{eq:spherical_Y_lm_ident3}\\
	\sin^3\theta Y^{lm}_{,\theta} &=& \zeta^{lm}_{(+3)} Y^{l+3,m} + \zeta^{lm}_{(+1)}Y^{l+1,m} + \zeta^{lm}_{(-1)}Y^{l-1,m} + \zeta^{lm}_{-3}Y^{l-3,m}\p \label{eq:spherical_Y_lm_ident6}
\end{eqnarray}
Defining
\begin{equation}
	C_{lm}=\left[\frac{l^2-m^2}{(2l+1)(2l-1)}\right]^{1/2}\c
\end{equation}
the explicit form of the $lm$-dependent coefficients ($\delta,\zeta$) in Eqs.~\eqref{eq:spherical_Y_lm_ident3} and \eqref{eq:spherical_Y_lm_ident6} are given by
\begin{gather}
	\delta^{lm}_{(+1)} = l C_{l+1,m}\c \qquad \delta^{lm}_{(-1)} = -(l+1)C_{lm}\c	\\
	\epsilon^{lm}_{(+1)} = (1-l)C_{l+1,m}\c\qquad \epsilon^{lm}_{(-1)} = (l+2)C_{lm}\c
\end{gather}
\begin{align}
	\zeta^{lm}_{(+3)} &= -l C_{l+1,m}C_{l+2,m}C_{l+3,m}		\c							\\
	\zeta^{lm}_{(+1)} &= C_{l+1,m}[l(l-C^2_{l+1,m} - C^2_{l+2,m}) + (l+1)C^2_{lm}]\c		\\
	\zeta^{lm}_{(-1)} &= -C_{lm}[(l+1)(1-C^2_{l-1,m}-C^2_{lm})+l C^2_{l+1,m}]\c	\\
	\zeta^{lm}_{(-3)} &= (l+1)C_{lm}C_{l-1,m}C_{l-2,m}\p
\end{align}

\bibliography{references}

\end{document}